\documentclass[prl, twocolumn, superscriptaddress, nofootinbib]{revtex4-2}

\usepackage[english]{babel}
\usepackage{hyperref}

\usepackage[strict]{changepage}

\usepackage[pdftex]{graphicx}
\usepackage{color}
\usepackage[toc,page]{appendix}

\usepackage{amsmath}
\usepackage{amssymb}
\usepackage{mathtools}
\usepackage{braket}
\usepackage{stmaryrd}

\usepackage[sort&compress]{natbib}

\newcommand{\un}[1]{\ensuremath{\,\mathrm{#1}}}
\renewcommand{\v}[1]{\ensuremath{\boldsymbol{#1}}}

\newcommand{\fig}[1]{Figure~\ref{fig:#1}}

\newcommand{\lr}[1]{\ensuremath{\left( #1 \right)}}

\newcommand{\I}{\mathrm{i}}

\newcommand{\mc}{\mathcal}

\newcommand{\abs}[1]{\left| #1 \right|}
\newcommand{\vkb}{\ensuremath{\v{k}_{\text{b}}}}
\newcommand{\vkt}{\ensuremath{\v{k}_{\text{t}}}}
\newcommand{\vrb}{\ensuremath{\v{r}_{\text{b}}}}
\newcommand{\vrt}{\ensuremath{\v{r}_{\text{t}}}}

\begin{document}

\title{Steering the current flow in twisted bilayer graphene}

\author{Jesús~Arturo~Sánchez-Sánchez}
\email{jasanchez@icf.unam.mx}
\affiliation{Instituto de Ciencias F\'isicas, Universidad Nacional Aut\'onoma de M\'exico, Cuernavaca, México}

\author{Montserrat~Navarro-Espino}
\email{monsene10@gmail.com}
\affiliation{Facultad de Química, Universidad Nacional Aut\'onoma de M\'exico, Ciudad de México, México}

\author{Yonatan~Betancur-Ocampo}
\email{ybetancur@icf.unam.mx}
\affiliation{Instituto de Ciencias F\'isicas, Universidad Nacional Aut\'onoma de M\'exico, Cuernavaca, México}

\author{José~Eduardo~Barrios-Vargas}
\email{j.e.barrios@gmail.com}
\affiliation{Facultad de Química, Universidad Nacional Aut\'onoma de M\'exico, Ciudad de México, México}

\author{Thomas~Stegmann}
\email{stegmann@icf.unam.mx}
\affiliation{Instituto de Ciencias F\'isicas, Universidad Nacional Aut\'onoma de M\'exico, Cuernavaca, México}

\date{\today}

\begin{abstract}
A nanoelectronic device made of twisted bilayer graphene (TBLG) is proposed to steer the direction of the current flow. The ballistic electron current, injected at one edge of the bottom layer, can be guided predominantly to one of the lateral edges of the top layer. The current is steered to the opposite lateral edge, if either the twist angle is reversed or the electrons are injected in the valence band instead of the conduction band, making it possible to control the current flow by electric gates. When both graphene layers are aligned, the current passes straight through the system without changing its initial direction. The observed steering angle exceeds well the twist angle and emerges for a broad range of experimentally accessible parameters. It is explained by the trigonal shape of the energy bands beyond the van Hove singularity due to the Moir\'e interference pattern. As the shape of the energy bands depends on the valley degree of freedom, the steered current is partially valley polarized. Our findings show how to control and manipulate the current flow in TBLG. Technologically, they are of relevance for applications in twistronics and valleytronics. 
\end{abstract}

\maketitle

\section{Introduction}

Since its first experimental isolation almost two decades ago \cite{Novoselov2004}, graphene is surprising continuously with exceptional properties, making it probably the most studied material in the world with several potential device applications, see \cite{Foa-Torres2020} for an introduction and \cite{Celasco2019} for a detailed overview. One of the most prominent properties is certainly the Dirac behavior of the electrons, making graphene a condensed matter analog for phenomena from quantum electrodynamics \cite{Katsnelson2006, Novoselov2005, Katsnelson2007}. However, interesting characteristics are not limited to single layers of graphene but are found also in stacks of this material. For example, bilayer graphene (BG), while also being a gapless semiconductor, has a low-energy dispersion that is not linear but quadratic \cite{McCann2013} and features anti-Klein tunneling \cite{Katsnelson2006, Kleptsyn2015, Du2018}.

Graphene research has shown recently a sudden 'twist' with the observation of unconventional superconductivity and Mott-like insulating states in twisted bilayer graphene (TBLG) \cite{Cao2018-1, Cao2018-2}, which has stimulated an avalanche of work, see \cite{MacDonald2019, Catarina2019, Mogera2020, Andrei2020, Nimbalkar2020} for an overview, and eventually initiated a new research field called \textit{twistronics} \cite{Carr2017, Ribeiro-Palau2018}. The observed effects can be attributed generally to the emergence of flat bands at certain magic angles, the largest of them is slightly above 1.1 degree \cite{Tarnopolsky2019, Lisi2020}. These flat bands are manifested in the density of states as van Hove singularities at electron densities accessible by electrostatic gating. As the electrons in these bands are densely packed in momentum space, they become strongly correlated \cite{Jiang2019, Choi2019, Xie2019, Wong2020, Xu2021, Choi2021, Gadelha2021}. Moreover, the twist angle allows to tune the flatness of the bands and therefore, the correlations, making TBLG an ideal platform to study and emulate strongly correlated systems \cite{Kennes2021}. Some of the discovered insulating states even show ferromagnetism, implying that these states are topological \cite{Lu2019, Sharpe2019, Song2019, Nuckolls2020, Bultinck2020, Ma2020}. Below the magic angle, interesting effects like one-dimensional channels and electronic localization have been reported \cite{Dai2016, Walet2019, Fleischmann2019, Nguyen2021}.

While most of the research is dedicated to the correlations at the magic angles, the effects caused by the structural changes of the energy bands have received less attention. In this direction, it has been demonstrated that TBLG shows a topological phase transition \cite{Kim2016} and can be used as a valley filter, making it a candidate for applications in valleytronics \cite{Hou2020, Schaibley2016}. The electronic transport in small TBLG samples (size about $3 \un{nm} $) has been investigated, showing conductance oscillations as a function of the twist angle \cite{Han2020}.

In this paper, we investigate the electrical transport in TBLG devices with a size of about $100 \un{nm}$. The Green's function method is applied to the established tight-binding model of TBLG to calculate numerically the current flow. We find that the current, injected at one edge of the bottom layer and detected at the edges of the top layer, is guided predominantly to one of the lateral edges. The current is directed to the opposite lateral edge, if the twist angle is reversed or if the electrons are injected in graphene’s valence band instead of the conduction band. We perform band structure calculations showing that the observed steering of the current is due to the trigonal shape of the energy bands at electron densities above the van Hove singularity due to the Moir\'e interference pattern. Moreover, we demonstrate that the deviated current is partially valley polarized as the energy bands depend on the valley spin of the electrons. Our findings show how to control and manipulate the current flow in TBLG by means of the twist angle and electrostatic gating, paving the way to new nanoelectronic devices.

\section{Modeling twisted bilayer graphene}

\begin{figure}[t]
   \centering
   \includegraphics[width=\linewidth]{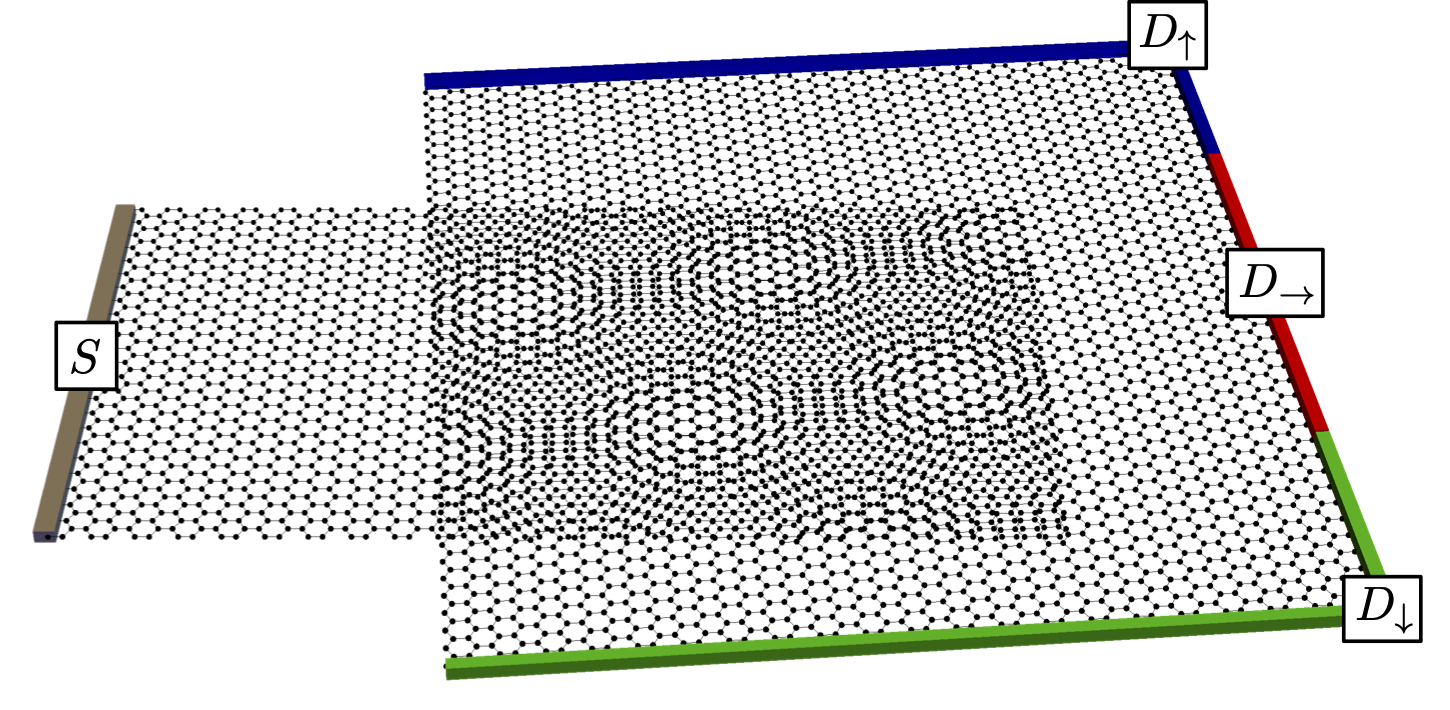}
   \caption{Schematic representation of the studied TBLG device. It consists of two stacked graphene nanoribbons, where the upper layer is twisted by the angle $\theta $ with respect to the lower one that remains fixed. Electrons are injected through the source contact $S$ at the left edge of the bottom layer (gray bar). They pass through the twisted bilayer region and are detected by three drain contacts $D_{\uparrow/\rightarrow/\downarrow}$ at the edges of the top layer (blue, red, and green bars).}
   \label{fig:1}
 \end{figure}

The studied device consists of two rectangular graphene nanoribbons that overlap in a certain region forming a graphene bilayer, see \fig{1}. The bottom and top layer, separated by a distance of $d_0=0.343\un{nm}$, have a size of $100 \times 60 \un{nm}$ and $86 \times 100 \un{nm}$ with a bilayer region of about $60 \times 60 \un{nm}$. The top layer can be twisted by the angle $\theta$ with respect to the bottom layer, which is kept fixed. In order to study the electrical transport through the system, a source contact $S$ injects electrons at the left edge of the bottom layer. The injected current passes through the twisted bilayer region and is detected by three drain contacts $D_{\uparrow/\rightarrow/\downarrow}$ at the edges of the top layer.

The TBLG device is modeled by the tight-binding Hamiltonian \cite{Catarina2019}
\begin{equation}
    \label{Htotal}
    H=H_{\text{b}}+H_{\text{t}}+H_{\perp}
\end{equation}
where $H_{\text{b/t}}$ corresponds to the isolated bottom/top graphene layer while $H_{\perp}$ takes into account the interlayer coupling. The bottom and top layer's Hamiltonians are the same as for pristine graphene 
\begin{equation}
    \label{H_SLG}
    H_{\text{b/t}}= t_{0}\sum_{(n,m)} \ket{n} \bra{m} + \text{h.c.} 
\end{equation}
where $\ket{n}$ indicates the atomic state localized on the carbon atom at $\v{r}_n$ on either sublattice A or B. The sum runs over the nearest neighboring carbon atoms separated by the distance $a_0=0.142\un{nm}$ and coupled with the energy $t_0=-2.7 \un{eV}$. The interlayer Hamiltonian reads
\begin{equation}
    \label{H_interlayer}
    H_{\perp}= \sum_{n,m} t_{nm}^{\perp} \ket{n_{\text{b}}} \bra{m_{\text{t}}} + \text{h.c.}
\end{equation}
where the sum is over pairs of atomic states that belong to the bottom and top layer. The interlayer coupling is determined by the Slater-Koster formula
\begin{equation}
    \label{t_perp}
    t_{nm}^{\perp}= \cos^{2} (\gamma) V_{pp\sigma}(r_{nm})+ [1-\cos^{2}(\gamma)]V_{pp\pi}(r_{nm})
\end{equation}
where $\cos\gamma=z_{nm}/r_{nm}$ is the direction cosine of $\v{r}_{nm}=\v{r}_m-\v{r}_n$ along the $z$ direction. The Slater-Koster parameters are defined as
\begin{equation}
    \label{Slater-Koster}
    \begin{split}
    V_{pp\sigma}(r_{nm})&=V_{pp\sigma}^{0}\exp\left[q_{\sigma} \Bigl (1-\frac{r_{nm}}{d_0} \Bigr)\right],\\
    V_{pp\pi}(r_{nm})&=V_{pp\pi}^{0}\exp\left[q_{\pi} \Bigl(1-\frac{r_{nm}}{a_0} \Bigr) \right].
    \end{split}
\end{equation}
with the tight-binding parameters \cite{Gadelha2021}
\begin{equation}
\label{TB_parameters}
    \begin{split}
    V_{pp\sigma}^{0}=0.375 \un{eV}, \quad V_{pp\pi}^{0}=-2.7 \un{eV},\\[1mm]
    \frac{q_{\pi}}{a_0}=\frac{q_{\sigma}}{d_0}=22.18 \un{nm}^{-1}. \quad
    \end{split}
\end{equation}

\subsection{Band structure calculations}

\begin{figure*}[t]
   \centering
   \includegraphics[width=\linewidth]{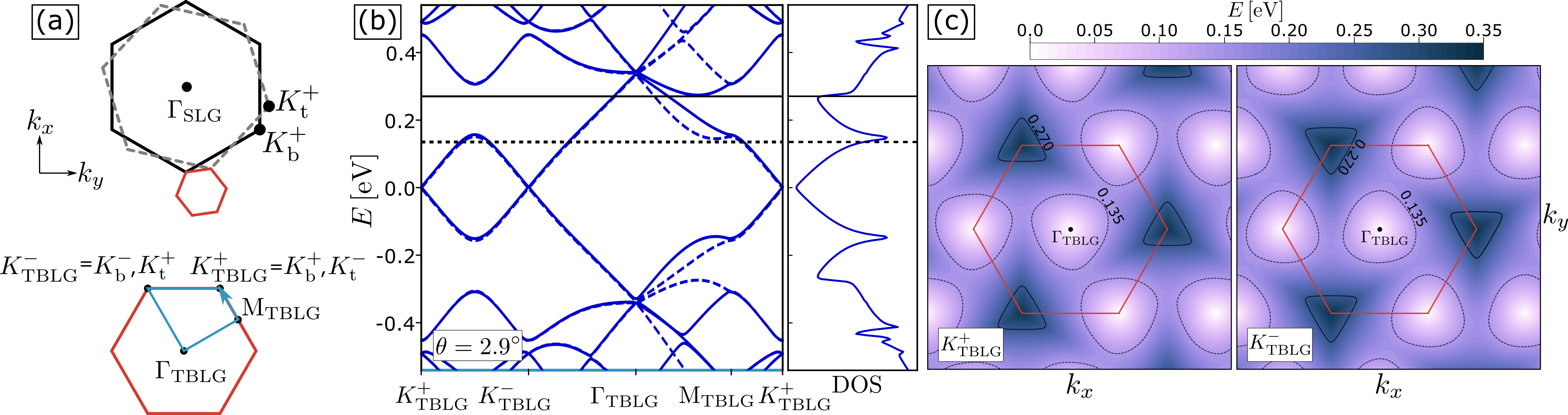}
   \caption{(a) The twist of the two monolayer Brillouin zones (black and gray dashed hexagons) generates the TBLG Brillouin zone (red hexagon). A zoom into this Brillouin zone shows the high symmetry points, in particular the $K^\pm_\text{TBLG}$ points and their relation to the monolayer. (b) Band structure along the blue path indicated in (a) and density of states (DOS) at a twist angle of $\theta=2.9^\circ$. (c) Constant energy contours of the band structure close to the two $K^\pm_\text{TBLG}$ points at $\theta=2.9^\circ$.}
   \label{fig:2}
 \end{figure*}

We model the band structure of TBLG using the low-energy continuum limit of \eqref{Htotal} for a single valley. This model describes small twist angles accurately ($\theta<10^\circ$). Each graphene layer is represented by a Dirac Hamiltonian
\begin{align}
    H_{\text{b/t}}^\pm = \hbar v_F \v{q} \cdot \left(\pm \sigma_x^{\theta_{\text{b/t}}}, \sigma_y^{\theta_{\text{b/t}}}\right),
\end{align}
where $\hbar v_F= 3 a_0 t_0/2$ and the $\pm$ signs correspond to the different valleys $K_\text{TBLG}^\pm$, see \fig{2}~(a). The twist of the graphene layers is taken into account within the vector
\begin{align}
    \begin{pmatrix}
        \sigma_x^{\theta_{\text{b/t}}} \\
        \sigma_y^{\theta_{\text{b/t}}} 
    \end{pmatrix}
    =
    \begin{pmatrix}
        \cos\theta_{\text{b/t}} & -\sin\theta_{\text{b/t}} \\
        \sin\theta_{\text{b/t}} & \cos\theta_{\text{b/t}}
    \end{pmatrix}
    \begin{pmatrix}
        \sigma_x \\
        \sigma_y
    \end{pmatrix},
\end{align}
which consists of rotated Pauli matrices with $\theta_\text{b}=0^\circ$ (as the bottom layer is fixed) and $\theta_\text{t}=\theta$. We transform the interlayer Hamiltonian \eqref{H_interlayer} to the reciprocal Fourier space
\begin{align}
    H_\perp= \sum_{\vkb, \vkt} \sum_{\alpha, \beta} \ket{\vkb, \alpha} t_{\alpha,\beta}^\perp(\vkb,\vkt) \bra{\vkt,\beta},
\end{align}
where (apart from a normalization factor)
\begin{align}
    t_{\alpha,\beta}^\perp(\vkb,\vkt) = 
    \int \frac{d^2\v{p}}{2\pi} \sum_{\vrb, \vrt} 
    e^{-\I(\vkb-\v{p})\cdot \vrb^\alpha}
    \, t^\perp(p) \,
    e^{\I(\vkt-\v{p})\cdot \vrt^\beta}.
\end{align}
The indices $\alpha$ and $\beta$ indicate the two different carbon atoms (A/B) in each unit cell with the positions $\vrb^\alpha= \vrb + \v\tau_\alpha$ and $\vrt^\beta= \vrt + \v\tau_\beta$, respectively. The function $t_\perp(p)$ is the Fourier transform of the interlayer coupling \eqref{t_perp}. Taking into account only the terms in leading order (using $t_\perp(|K_\text{TBLG}^\pm|)= 0.433\un{eV \AA^2}$), we obtain in this way the band structure shown in \fig{2}~(b) along the blue path in (a), while constant energy contours are depicted in (c). Further details on these calculations can be found in \cite{Catarina2019}.

\subsection{The Green's function method for electron transport}

The nonequilibrium Green's function (NEGF) method is applied to study the electrical transport in the TBLG device. As detailed introductions into this method can be found in various textbooks \cite{Datta1997, Datta2005, DiVentra2008}, we summarize here briefly the essential equations. The Green's function of the system is given by
\begin{equation}
    \label{GF}
    G=\Bigl(E-H-\sum_p\Sigma_p\Bigr)^{-1}
\end{equation}
where $E$ is the energy of the injected electrons and $H$ is the tight-binding Hamiltonian \eqref{Htotal}. The contacts are modeled by the so-called wide-band model implying for these a constant energy-independent surface density of states \cite{Verzijl2013}. It is taken into account in the Green's function by means of the self-energies
\begin{equation}
    \label{Self-E}
    \Sigma_{p}= -\I \sum_{n \in C_{p}} \ket{n}\bra{n}
\end{equation}
where the sum runs over the carbon atoms $C_p$ attached to the contact $ p \in \{S,D_{\uparrow, \rightarrow, \downarrow}\} $, see the colored bars in \fig{1}.

The transmission between a pair of contacts, $i$ and $j$, is given by
\begin{equation}
    \label{Transmission}
	T_{ij}(E)= \mathrm{Tr}[\Gamma_{i}G\Gamma_{j}G^{\dagger}],
\end{equation}
where the inscattering function associated to contact $i$ is defined as $\Gamma_{i}= -2\mathrm{Im}(\Sigma_i)$. The transmission determines the conductance for electrons of energy $E$ between the selected pair of contacts, $\mathcal{G}_{ij}= e^2/h \, T_{ij}$. Finally, the local current flowing between the atoms $n$ and $m$ can be calculated by
\begin{equation}
    \label{Current}
    I_{nm}(E)= \mathrm{Im}\bigl[t^{*}_{nm} (G\Gamma_{S}G^{\dagger})_{nm} \bigr].
\end{equation}

\begin{figure*}
   \centering
   \includegraphics[width=1.0\linewidth]{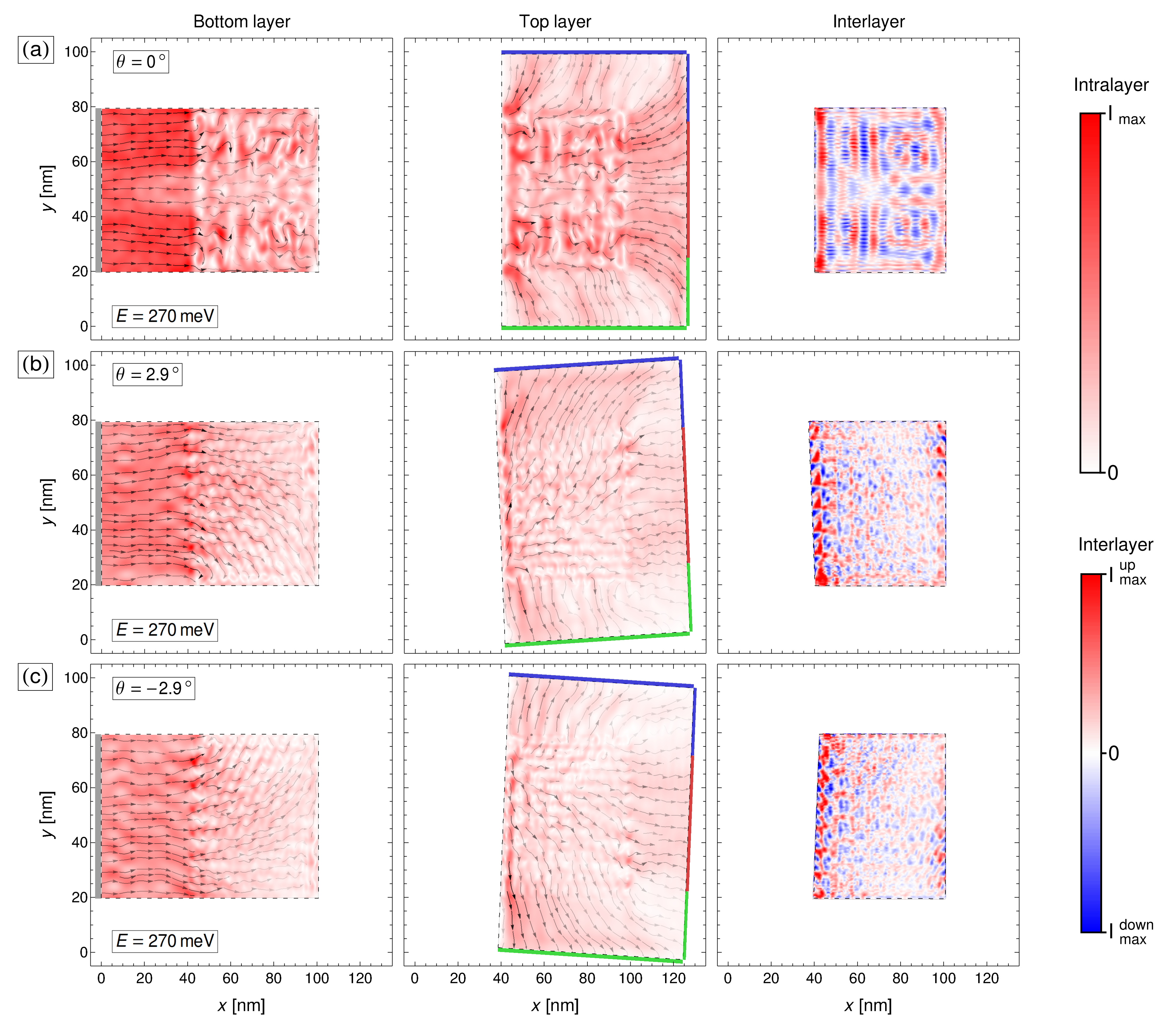}\\[-3mm]
   \caption{Current flow in the TBLG device. The intralayer current (left and central column) is indicated by the black arrows and its density by the red color shading. The interlayer current density (right column) that goes from the bottom to the top layer is given by reddish colors, while for the opposite direction bluish colors are used. When both layers are aligned (row a), the current goes essentially straight through the system. When a positive twist angle is introduced (row b), the current density is steered to the upper drain contact $D_\uparrow$ at the top layer, while for a negative twist (row c) the current is guided to $D_\downarrow$.}
   \label{fig:3}
\end{figure*}

\subsection{Calculating the longitudinal and the Hall resistance}

The longitudinal and the Hall resistance are defined as 
\begin{equation}
\label{Rxx-Rxy}
    \mc{R}_{xx}= \frac{V_{D_\rightarrow}-V_S}{I_{SD_\rightarrow}},
    \qquad    
    \mc{R}_{xy}= \frac{V_{D_\uparrow}-V_{D_\downarrow}}{I_{SD_\rightarrow}},
\end{equation}
where the $V_i$ are the potentials of the four contacts and $I_{SD_\rightarrow}$ is the current flowing between $S$ and $D_\rightarrow$. To calculate these quantities within the Green's function approach, we apply the Landauer-B\"uttiker equation for the total current at the $j$th contact \cite{Datta1997, Datta2005, DiVentra2008}
\begin{equation}
    \label{LBeq}
    I_j(E)= \frac{e}{h} \, \sum_{i} T_{ij} \lr{V_i-V_j},
\end{equation}
and use the fact that the current is zero at the contacts $D_{\uparrow,\downarrow}$ of unknown potential (due to the infinite internal resistance of a voltmeter). After some straightforward matrix algebra, we obtain finally \cite{StegmannPhd2014, Stegmann2015}
\begin{align}
    \label{Rxx-Rxy-Green}
    \mc{R}_{xx}&= \frac{h}{e^2} \, \frac{1}{T_{SD_\rightarrow} + \sum_{ij} T_{Si} \mc{M}_{ij} T_{jD_\rightarrow}}, \notag \\[2mm]
    \mc{R}_{xy}&= \frac{h}{e^2} \frac{\sum_j T_{Sj} \lr{\mc{M}_{jD_\uparrow}-\mc{M}_{j D_\downarrow}}}{T_{SD_\rightarrow} + \sum_{ij} T_{Si} \mc{M}_{ij} T_{jD_\rightarrow}}
\end{align}
where
\begin{equation}
    \label{Mmat}
    \mc{M}^{-1}=
   \begin{cases}
   -T_{ij} \quad &\text{if} \quad i\neq j,\\
   \sum_{k\neq i} T_{ik} \quad &\text{if} \quad i=j.
   \end{cases}
\end{equation}
The sums in \eqref{Rxx-Rxy-Green} are over the contacts $D_{\uparrow,\downarrow}$ of unknown potential, whereas the sums in \eqref{LBeq} and \eqref{Mmat} are over all contacts in the device.

\begin{figure*}[t]
  \centering
  \includegraphics[width=\linewidth]{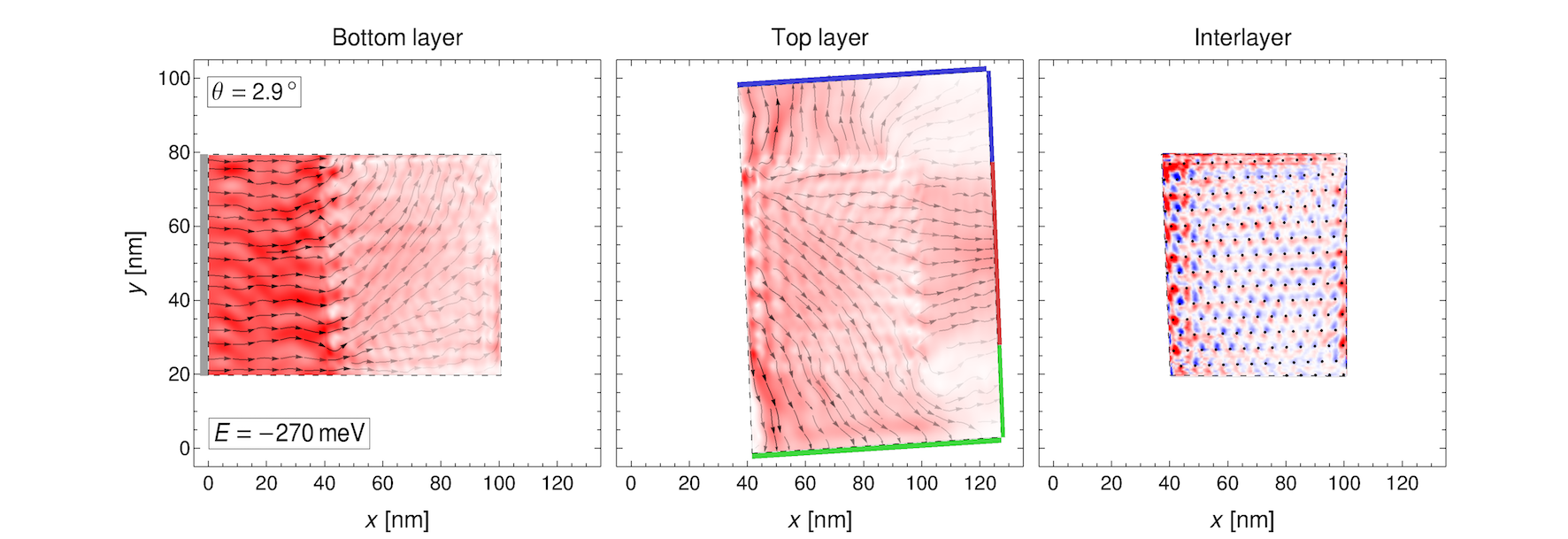}
  \caption{Current flow in the TBLG device at electron energy $E=-270 \un{meV}$ and a twist angle of $\theta = 2.9^\circ$. The steering of the current flow can be controlled by the electron energy, accessible experimentally through electrostatic gating. The color scaling is the same as in \fig{3}. The black dots in the interlayer current indicate the regions of AA stacking symmetry.}
  \label{fig:4}
\end{figure*}

\begin{figure}[t]
  \centering
  \includegraphics[width=0.8\linewidth]{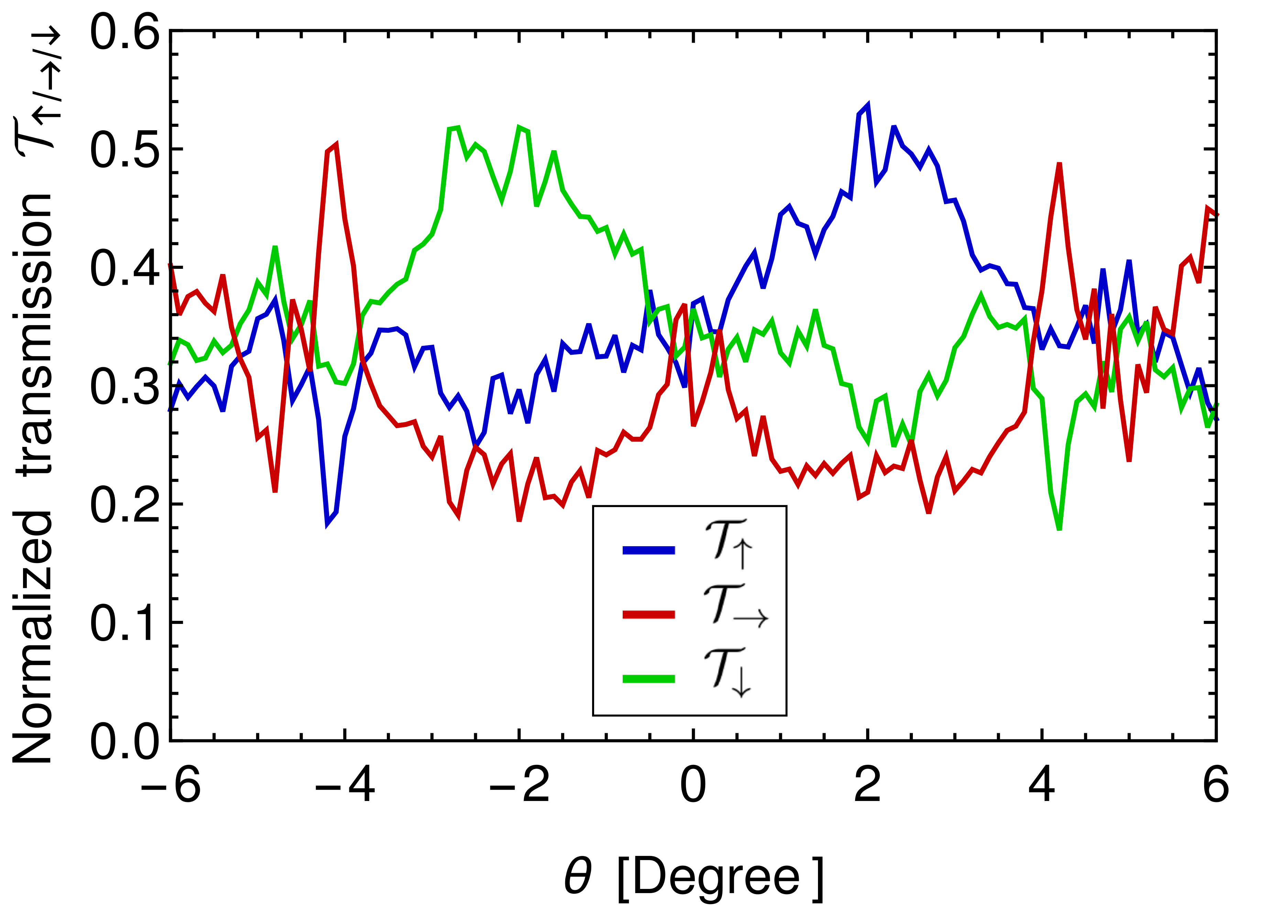}
  \caption{Normalized transmission $\mc{T}_{\uparrow/\rightarrow/\downarrow}$ for electrons at energy $E=270 \un{meV}$ as a function of the twist angle. For $1^\circ \lesssim \theta \lesssim 5^\circ $ a significantly larger fraction of the current is directed to contact $D_\uparrow$. When the twist angle is reversed the larger part of the current goes to contact $D_\downarrow$.}
  \label{fig:5}
\end{figure}

\subsection{Calculating the valley polarization}

\begin{figure*}[t]
   \centering
   \includegraphics[width=\linewidth]{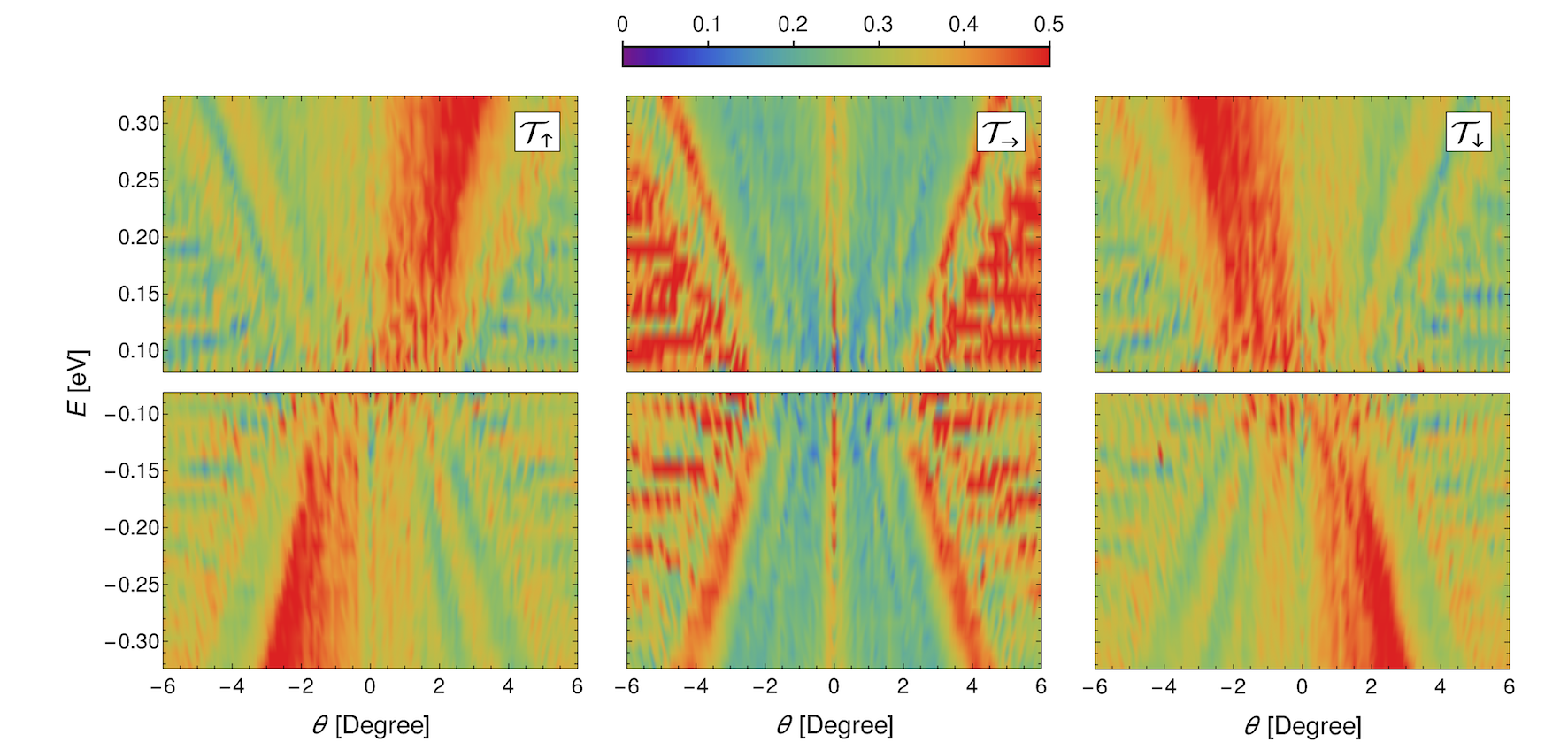}
   \caption{Normalized transmissions $\mc{T}_{\uparrow/\rightarrow/\downarrow}$ as a function of the twist angle and the electron energy. The broad red regions that are present in $\mc{T}_{\uparrow}$ but absent in $\mc{T}_\downarrow$ (or vice versa) prove the steering of the current flow.}
   \label{fig:6}
\end{figure*}

\begin{figure}[t]
    \centering
    \includegraphics[width=\linewidth]{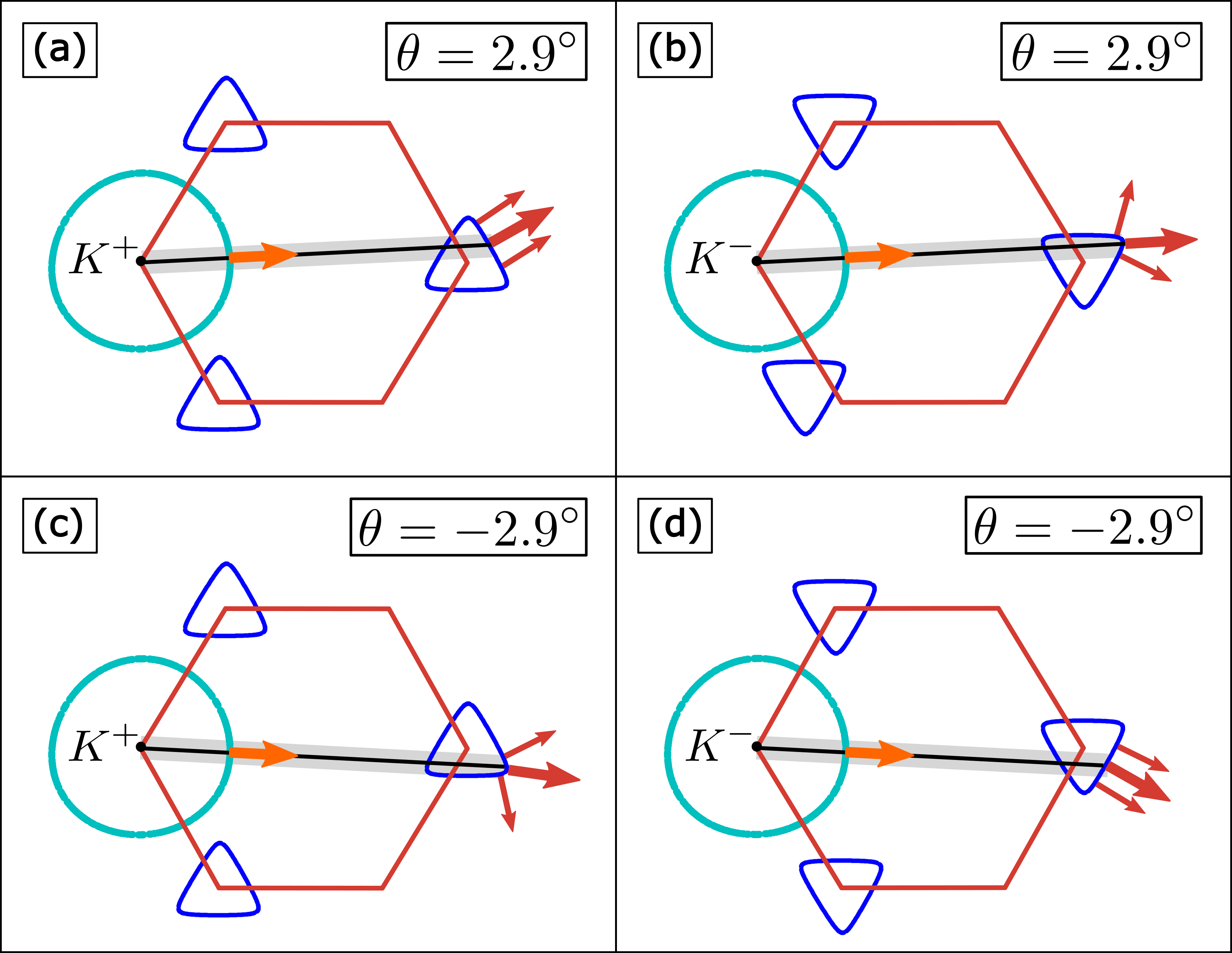}
    \caption{Calculated constant energy contours at $E=270\un{meV}$ and kinematical construction to understand the steering of the current flow. The trigonal shape of the energy contours in the bilayer region (blue triangles) changes the direction of the incident electron beam, indicated by the orange arrow perpendicular on the monolayer isoline (turquoise circle), in a preferential direction depending on the twist angle and the valley. (a,b) Energy contours at the $K^\pm$ valley for a twist angle of $\theta= 2.9^\circ$. (c,d) Contours for a twist of $\theta=-2.9^\circ$.}
   \label{fig:7}
\end{figure}

The valley polarization of a state $\ket{\phi}$ characterizes to which degree this state occupies the two valleys $K^+$ and $K^-$ in graphene. It can be calculated by the projection $\mc{P}(\v k)= \abs{\braket{\psi(\v k)| \phi}}^2$ onto the graphene eigenfunctions 
\begin{equation}
    \label{Psi_n}
    \psi_{n}(\v{k})=
    \begin{cases}
        e^{\I\v{k}\cdot\v{r}_n}, &n \in A\\
        \text{sign}(E)\, e^{\I\v{k}\cdot(\v{r}_n-\v{\delta})}\frac{f(\v{k})}{|f(\v{k})|}, &n \in B
    \end{cases}
\end{equation}
where
\begin{equation}
    \label{f_k}
    f(\v{k})=-t_0  \bigl[ e^{-\I \v{k} \cdot \v a_1} +e^{-\I \v{k} \cdot \v a_2} +e^{-\I \v{k} \cdot (\v a_2 + \v a_2)} \bigr]
\end{equation}
and $\v{a}_{1/2}= (3,\pm\sqrt{3})a_0/2$ are the lattice vectors while $\v\delta = (1,0)a_0$ is the vector connecting the sublattices A and B. Within the Green's function approach, this projection reads \cite{Stegmann2019, Settnes2016}
\begin{equation}
    \label{P_i_k}
    P_{i}(\v k)=\braket{\psi(\v{k})|G^\text{in}|\psi(\v{k})}_{A_i}.
\end{equation}
It is calculated over a finite region $A_i$ of the system, see for example the gray-shaded regions in \fig{8} (top right), in order to assess the valley polarization locally. The spectral density $P_i(\v k)$ is integrated around the valleys
\begin{equation}
    \label{P_+-}
    \mc{P}_{i}^{\pm}=\int_{\v{k}\in K^{\pm}}d^2k \, P_{i}(\v{k})
\end{equation}
and the valley polarization is given by
\begin{equation}
    \label{Pi}
    \mc{P}_{i}=\frac{\mc P_{i}^{+}-\mc P_{i}^{-}}{\mc P_{i}^{+}+\mc P_{i}^{-}}.
\end{equation}
For $\mc P_i = \pm 1$ the electrons are localized exclusively at the $K^\pm$ valleys and hence, are completely valley polarized, whereas for $\mc P_i=0$ they are completely unpolarized.

\section{Current flow in TBLG}

\begin{figure*}[t]
   \centering
   \includegraphics[width=0.95\linewidth]{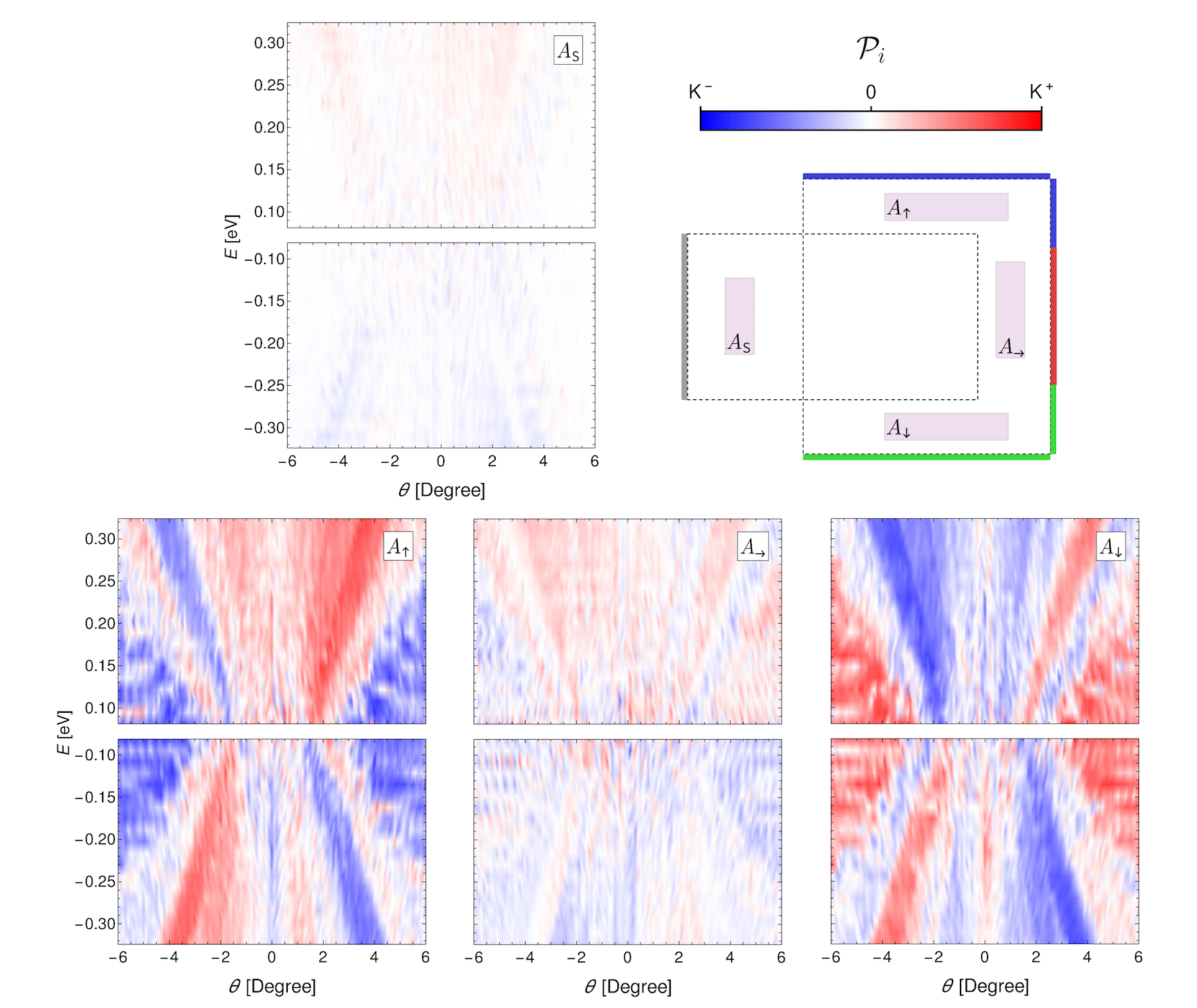}\\[-3mm]
   \caption{Valley polarization of the current in the TBLG device, calculated in the shaded rectangular regions as indicated in the device sketch. The current, injected unpolarized (region $A_S$), is valley polarized in the lateral regions $A_{\uparrow, \downarrow}$, whereas it remains unpolarized in $A_\rightarrow$.}
   \label{fig:8}
\end{figure*}

We start our discussion by analyzing in \fig{3} the local current flow in the TBLG device. In the absence of any twist (row a), we observe that the electrons, injected at energy $E=270\un{meV}$ by the contact $S$ at the left edge of the bottom layer, flow essentially straight through the system towards the right edge of the top layer. Minor deviations to the lateral edges are observed, which can be attributed to both, the diffraction of the electrons and boundary effects. However, these currents go in approximately equal amounts to the upper and lower edge of the top layer. The interlayer current\footnote{As the interlayer current shows some rare peaks, its maximum value is defined here as three standard deviations above the mean.} shows that at the edges of the bilayer region, the current is going from the bottom to the top layer (as expected due to the position of the contacts), while inside the bilayer region, it is oscillating between the layers.

When a twist angle of $\theta= 2.9^\circ$ is introduced (row b), the current changes its direction in the bilayer region. At the bottom of the bilayer region, the current is directed downwards, while at the top, it is propagating upwards. Eventually, the largest part of the current is directed to the drain contact $D_\uparrow$ at the upper edge of the top layer, while less current flows to the right and the bottom contacts, $D_\rightarrow$ and $D_\downarrow$. When the sign of the twist angleis reversed, $\theta= -2.9^\circ$ (row c), the current pattern is reversed as well and the main part of the current is directed to $D_\downarrow$. 

The steering of the current flow can also be reversed by injecting the electrons in the valence band instead of the conduction band, see \fig{4}. Similar to the case of aligned layers, the interlayer current shows that a large part of the current is passing to the top layer at the edges of the bilayer region, while it oscillates strongly inside the bilayer region. These oscillations are correlated with the regions of AA stacking symmetry, indicated by small black dots. The steering of the electrical transport in TBLG, observed qualitatively in the current flow patterns in \fig{3} and \fig{4}, is the main finding of this paper and will be analyzed in the following in detail.

In order to quantify our observations, we show in \fig{5} the normalized transmission
\begin{equation}
    \mc T_{\uparrow / \rightarrow / \downarrow}= \frac{T_{SD_{\uparrow / \rightarrow / \downarrow}}}{T_{SD_\uparrow}+T_{SD_\rightarrow}+T_{SD_\downarrow}}
\end{equation}
between the source and the three drain contacts. For a broad range of twist angles, roughly between 1 and 5 degree, a significantly larger fraction of the current is guided towards the contact $D_\uparrow$ as to the contacts $D_\rightarrow$ or $D_\downarrow$. For negative twist angles, the situation is reversed and the larger part of the current goes to the contact $D_\downarrow$. Note that the transmission curves are in good approximation symmetric, $\mc{T}_{\uparrow/\downarrow}(\theta) \simeq \mc{T}_{\downarrow/\uparrow}(-\theta)$, which can be expected as the overall structure of the Moir\'e interference pattern\footnote{In the following we neglect the subtle detail that some twist angles do not generate a Moir\'e pattern.} of the two layers is the same for positive and negative twist angles. However, due to the finite size of the investigated device, the precise position of the pattern can change (for example, the position of the AA regions) leading to the observed minor deviations from the symmetry of the transmission curves.

\begin{figure*}[t]
  \centering
  \includegraphics[width=\linewidth]{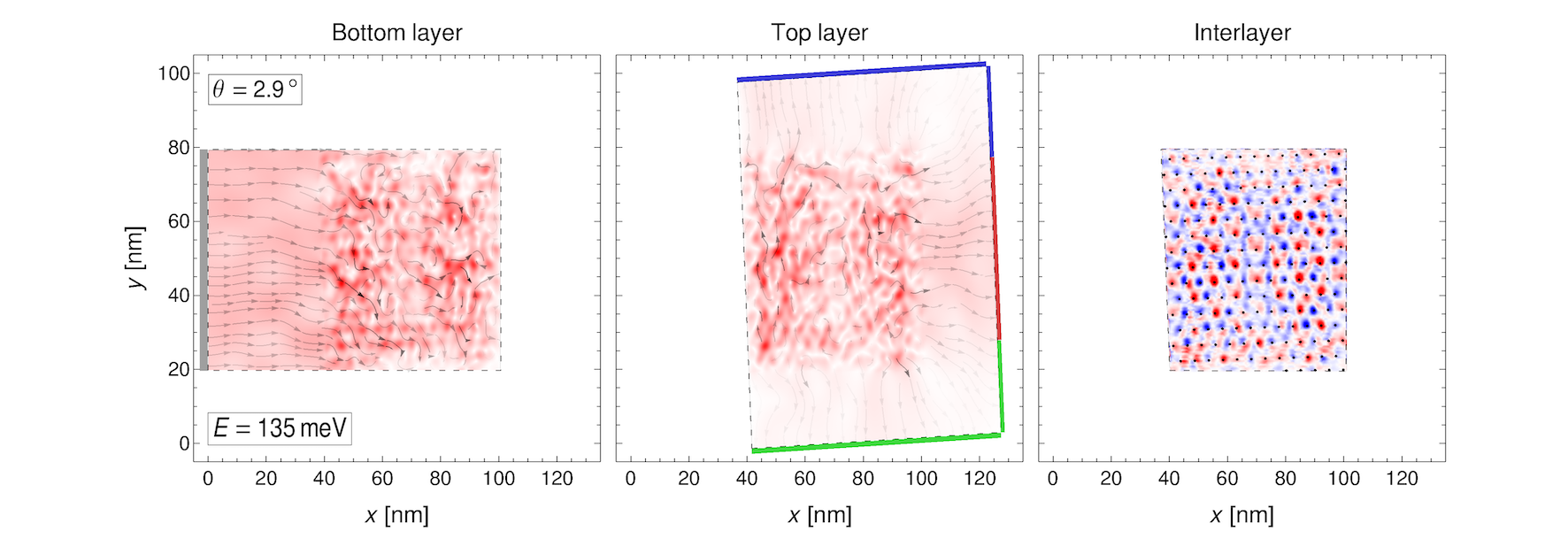}\\[-2mm]
  \caption{Current flow in the TBLG device at electron energy $E=135 \un{meV}$ and a twist of $\theta = 2.9^\circ$. Steering of the current flow to one of the lateral edges cannot be observed as the energy contours for these parameters are no longer triangular.}
  \label{fig:9}
\end{figure*}

\begin{figure}[t]
   \centering
   \includegraphics[width=\linewidth]{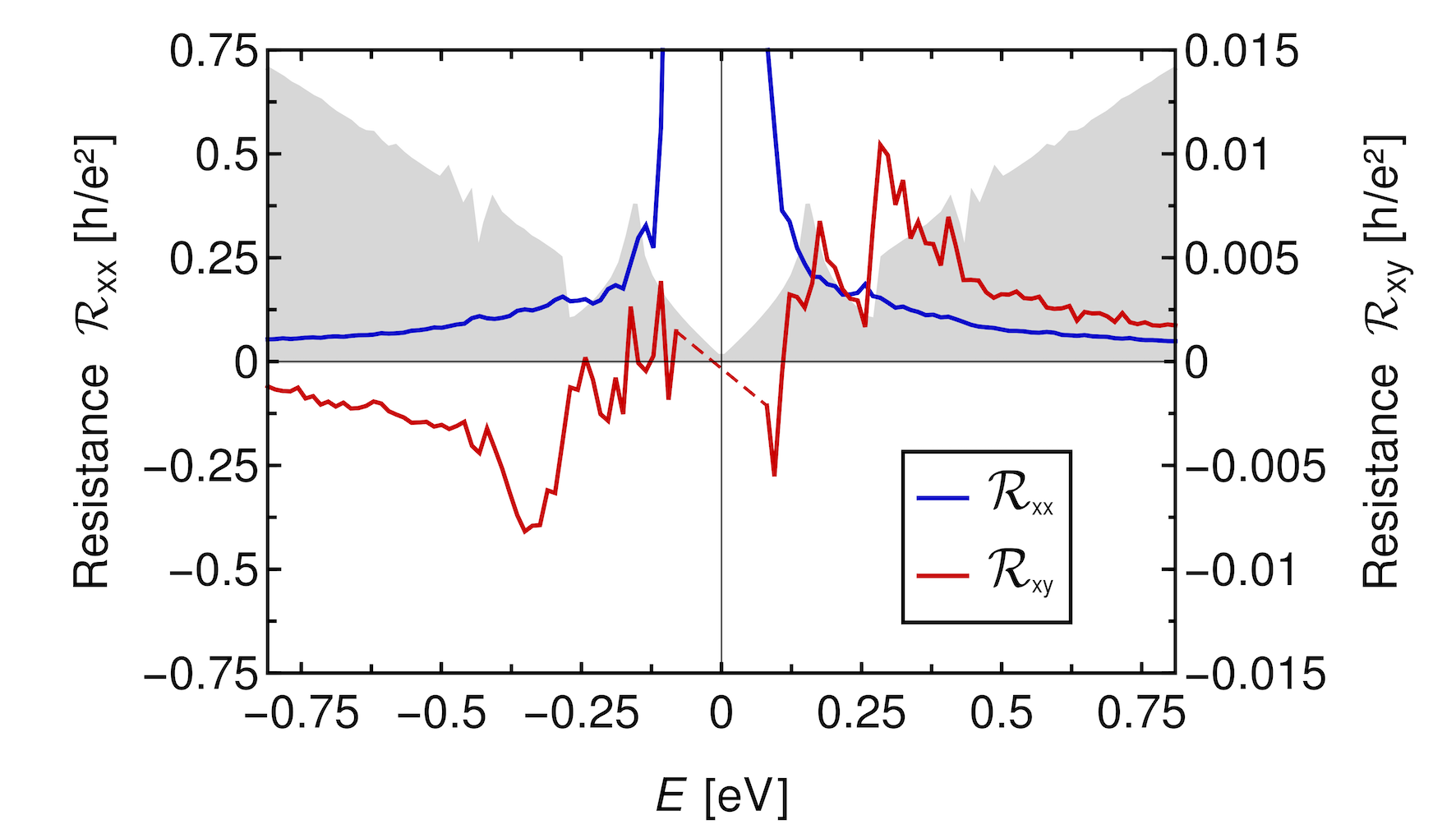}
   \caption{Longitudinal resistance $\mc{R}_{xx}$ (blue curve) and Hall resistance $\mc{R}_{xy}$ (red curve) as a function of energy for the TBLG device at a twist angle of $\theta = 2.9^\circ$. The steering of the current flow to one of the lateral edges generates a non-local Hall resistance. The DOS (in arbitrary units) is indicated by the gray color shading.}
   \label{fig:10}
\end{figure}

In \fig{6} we show the normalized transmissions to the three drain contacts as a function of the twist angle and the electron energy. The broad red colored regions in $\mc{T}_{\uparrow/\downarrow}(\theta,E)$ confirm that for a wide range of (experimentally accessible) parameters a significantly larger part of the electric current is steered towards one of the lateral contacts $D_{\uparrow/\downarrow}$. The red region in $\mc{T}_\rightarrow$ indicates the parameter region, where the current passes essentially straight through the system without a significant change of the initial direction. The apparent symmetries in the positions of the red regions highlight how the steering of the current flow can be controlled by the twist angle and the electron energy (or electron density) that is typically accessible through electric gates.

Having demonstrated the steering of the current flow in the TBLG device, we are left to understand this behavior. This can be done by analyzing the constant energy contours of the bandstructure, see \fig{2}~(c), which have a trigonal shape within a certain energy range. In \fig{7}~(a), we show the energy contours at $E=270 \un{meV}$ around the $K^+$ valley within the monolayer region (turquoise circle) and the twisted bilayer region (blue triangles) for a twist angle of $\theta= 2.9^\circ$. The directional change of the current can be understood by the kinematical construction \cite{Garcia-Pomar2008, Betancur-Ocampo2018, Betancur-Ocampo2019}. The injected electron beam is indicated by the orange arrow perpendicular to the monolayer contour. The black line, with an inclination of $\theta$, represents the conservation of the momentum component parallel to the boundary of the two regions.\footnote{Note that instead of using a twisted line connecting the energy contours in the two regions, one could rotate also the contours in the bilayer region, see for example the small red hexagon in \fig{2}~(a).} Its intersection point with the energy contour in the bilayer region determines the direction of the current flow in that region, see the red arrow perpendicular to the bilayer contour. The direction of the current flow, which is initially aligned approximately parallel to the $k_x$ axis, is rotated upwards, due to the trigonal shape of the energy contour and the twist angle. The diffraction of the injected electron beam leads to non-zero $k_y$ values, which are taken into account by the opaque rectangle centered around the black line. Its intersection points with the TBLG contours lead to the small red arrows that still point all towards $D_\uparrow$.

In \fig{7}~(b), we repeat the above analysis for electrons in the $K^-$ valley, where the trigonal energy contours are rotated by $180^\circ$ compared to the $K^+$ valley. In this case, we find that the current is dispersed strongly to all three drain contacts. As the injected electrons occupy equivalently the two valleys, the kinematical construction proves that the larger part of the current is steered upwards. When the twist angle is reversed to $\theta= -2.9^\circ$ (or, equivalently, the electron energy is changed to $E= -270\un{meV}$), the triangular contours are rotated by $180^\circ$ and the current is directed predominantly to the bottom, see \fig{7}~(c,d).

Our analysis shows furthermore that the electrons in different valleys are steered in different directions, and therefore, the current flow to the lateral edges is partially valley polarized. To confirm this, we show in \fig{8} the valley polarization, calculated by means of \eqref{Pi} in specific regions of the device, see the shaded rectangles. The current close to the injecting source contact (region $A_S$) and the current passing straight through the device (region $A_\rightarrow$) are largely unpolarized. In contrast, the current that is directed to the lateral edges is valley polarized in opposite valleys, see the regions $A_{\uparrow}$ and  $A_{\downarrow}$.

As the steering of the current flow is due to the trigonal shape of the energy contours, it disappears when the contours change towards the isotropic shape. This is the case, for instance, when the energy of the electrons is lowered below the van Hove singularity of the Moir\'e interference pattern, see the current flow in \fig{9} at $E=135 \un{meV}$ and \fig{2}~(b-c).

Finally, we show in \fig{10} the longitudinal and Hall resistance of the device, quantities that are commonly accessible in experiments. The steering of the current flow towards one of the lateral edges leads to a non-zero Hall resistance. The DOS, indicated by the gray color shading, shows features similar to the Hall resistance and highlights implicitly the energy bands causing the current steering. The red dashed line indicates the region where the Hall resistance has not been calculated, because in this region the Fermi wavelength is larger than the system size and the calculation would suffer from finite-size effects. Our findings of a non-local resistance are loosely related to those reported recently by Ma and coworkers \cite{Ma2020}. In that work, the non-local resistance is generated by a topological state in the superconducting gap, while here the trigonal shape of the energy contours generates the non-local signal in a different parameter regime.

\section{Conclusions}

We have shown that the current flow in TBLG can be steered by the twist angle. Even more important, the direction is reversed if the electrons are injected in the valence band instead of the conduction band (or vice versa), making it possible to control the current flow in TBLG devices through electric gates. This directional control of the current flow in TBLG is observed for a broad range of experimentally accessible parameters, see \fig{6}.

We have explained the steering by the trigonal shape of the energy contours in the bilayer region using the kinematical construction, which also proved that the deviated currents are partially valley polarized, see \fig{8}. However, as this construction is based on a continuum model of the bilayer, it is naturally not able to explain why the currents in the bottom and top layer of the bilayer region flow in almost orthogonal directions, see \fig{3}. A splitting of the current flow due to the trigonal warping of graphene's energy bands has been reported before at the van Hove singularity of monolayer graphene \cite{Garcia-Pomar2008, Stegmann2019}. However, the steering in TBLG appears at experimentally accessible electron energies and can be controlled by the twist angle and electric gating. 

The proposed device allows to control and manipulate the current flow in TBLG. Technologically, it is of relevance for nanoelectronic applications and will ramp the rising field of twistronics. At this point, we can only speculate but the possibility to tune the direction of the current flow by gates could be used to construct a new type of graphene twist-transistor. Moreover, the fact that the steered current is partially valley polarized may be used to process not only the electron charge but also its valley degree of freedom.

\section{Acknowledgments}
JAS-S gratefully acknowledges a CONACYT graduate scholarship. MN-E acknowledges support from Subprograma 127-FQ, UNAM. We gratefully acknowledge funding from CONACYT Proyecto A1-S-13469, CONACYT Frontera 428214, UNAM-PAPIIT IA103020 and UNAM-PAPIIT IA103419.

\vspace*{-5mm}
\enlargethispage{10mm}

\bibliography{tblg-short}

\begin{thebibliography}{53}%
\makeatletter
\providecommand \@ifxundefined [1]{%
 \@ifx{#1\undefined}
}%
\providecommand \@ifnum [1]{%
 \ifnum #1\expandafter \@firstoftwo
 \else \expandafter \@secondoftwo
 \fi
}%
\providecommand \@ifx [1]{%
 \ifx #1\expandafter \@firstoftwo
 \else \expandafter \@secondoftwo
 \fi
}%
\providecommand \natexlab [1]{#1}%
\providecommand \enquote  [1]{``#1''}%
\providecommand \bibnamefont  [1]{#1}%
\providecommand \bibfnamefont [1]{#1}%
\providecommand \citenamefont [1]{#1}%
\providecommand \href@noop [0]{\@secondoftwo}%
\providecommand \href [0]{\begingroup \@sanitize@url \@href}%
\providecommand \@href[1]{\@@startlink{#1}\@@href}%
\providecommand \@@href[1]{\endgroup#1\@@endlink}%
\providecommand \@sanitize@url [0]{\catcode `\\12\catcode `\$12\catcode
  `\&12\catcode `\#12\catcode `\^12\catcode `\_12\catcode `\%12\relax}%
\providecommand \@@startlink[1]{}%
\providecommand \@@endlink[0]{}%
\providecommand \url  [0]{\begingroup\@sanitize@url \@url }%
\providecommand \@url [1]{\endgroup\@href {#1}{\urlprefix }}%
\providecommand \urlprefix  [0]{URL }%
\providecommand \Eprint [0]{\href }%
\providecommand \doibase [0]{https://doi.org/}%
\providecommand \selectlanguage [0]{\@gobble}%
\providecommand \bibinfo  [0]{\@secondoftwo}%
\providecommand \bibfield  [0]{\@secondoftwo}%
\providecommand \translation [1]{[#1]}%
\providecommand \BibitemOpen [0]{}%
\providecommand \bibitemStop [0]{}%
\providecommand \bibitemNoStop [0]{.\EOS\space}%
\providecommand \EOS [0]{\spacefactor3000\relax}%
\providecommand \BibitemShut  [1]{\csname bibitem#1\endcsname}%
\let\auto@bib@innerbib\@empty
\bibitem [{\citenamefont {Novoselov}\ \emph {et~al.}(2004)\citenamefont
  {Novoselov}, \citenamefont {Geim}, \citenamefont {Morozov}, \citenamefont
  {Jiang}, \citenamefont {Zhang}, \citenamefont {Dubonos}, \citenamefont
  {Grigorieva},\ and\ \citenamefont {Firsov}}]{Novoselov2004}%
  \BibitemOpen
  \bibfield  {author} {\bibinfo {author} {\bibfnamefont {K.~S.}\ \bibnamefont
  {Novoselov}}, \bibinfo {author} {\bibfnamefont {A.~K.}\ \bibnamefont {Geim}},
  \bibinfo {author} {\bibfnamefont {S.~V.}\ \bibnamefont {Morozov}}, \bibinfo
  {author} {\bibfnamefont {D.}~\bibnamefont {Jiang}}, \bibinfo {author}
  {\bibfnamefont {Y.}~\bibnamefont {Zhang}}, \bibinfo {author} {\bibfnamefont
  {S.~V.}\ \bibnamefont {Dubonos}}, \bibinfo {author} {\bibfnamefont {I.~V.}\
  \bibnamefont {Grigorieva}},\ and\ \bibinfo {author} {\bibfnamefont {A.~A.}\
  \bibnamefont {Firsov}},\ }\bibfield  {title} {\bibinfo {title} {Electric
  field effect in atomically thin carbon films},\ }\href
  {https://doi.org/10.1126/science.1102896} {\bibfield  {journal} {\bibinfo
  {journal} {Science}\ }\textbf {\bibinfo {volume} {306}},\ \bibinfo {pages}
  {666} (\bibinfo {year} {2004})}\BibitemShut {NoStop}%
\bibitem [{\citenamefont {Foa-Torres}\ \emph {et~al.}(2020)\citenamefont
  {Foa-Torres}, \citenamefont {Roche},\ and\ \citenamefont
  {Charlier}}]{Foa-Torres2020}%
  \BibitemOpen
  \bibfield  {author} {\bibinfo {author} {\bibfnamefont {L.~E.~F.}\
  \bibnamefont {Foa-Torres}}, \bibinfo {author} {\bibfnamefont
  {S.}~\bibnamefont {Roche}},\ and\ \bibinfo {author} {\bibfnamefont {J.-C.}\
  \bibnamefont {Charlier}},\ }\href {https://doi.org/10.1017/9781108664462}
  {\emph {\bibinfo {title} {Introduction to Graphene-Based Nanomaterials}}}\
  (\bibinfo  {publisher} {Cambridge University Press},\ \bibinfo {year}
  {2020})\BibitemShut {NoStop}%
\bibitem [{\citenamefont {Celasco}\ \emph {et~al.}(2019)\citenamefont
  {Celasco}, \citenamefont {Chaika}, \citenamefont {Stauber}, \citenamefont
  {Zhang}, \citenamefont {Ozkan}, \citenamefont {Ozkan}, \citenamefont {Ozkan},
  \citenamefont {Palys},\ and\ \citenamefont {Harun}}]{Celasco2019}%
  \BibitemOpen
  \bibinfo {editor} {\bibfnamefont {E.}~\bibnamefont {Celasco}}, \bibinfo
  {editor} {\bibfnamefont {A.~N.}\ \bibnamefont {Chaika}}, \bibinfo {editor}
  {\bibfnamefont {T.}~\bibnamefont {Stauber}}, \bibinfo {editor} {\bibfnamefont
  {M.}~\bibnamefont {Zhang}}, \bibinfo {editor} {\bibfnamefont
  {C.}~\bibnamefont {Ozkan}}, \bibinfo {editor} {\bibfnamefont
  {C.}~\bibnamefont {Ozkan}}, \bibinfo {editor} {\bibfnamefont
  {U.}~\bibnamefont {Ozkan}}, \bibinfo {editor} {\bibfnamefont
  {B.}~\bibnamefont {Palys}},\ and\ \bibinfo {editor} {\bibfnamefont {S.~W.}\
  \bibnamefont {Harun}},\ eds.,\ \href {https://doi.org/10.1002/9781119468455}
  {\emph {\bibinfo {title} {Handbook of Graphene}}}\ (\bibinfo  {publisher}
  {Wiley},\ \bibinfo {year} {2019})\BibitemShut {NoStop}%
\bibitem [{\citenamefont {Katsnelson}\ \emph {et~al.}(2006)\citenamefont
  {Katsnelson}, \citenamefont {Novoselov},\ and\ \citenamefont
  {Geim}}]{Katsnelson2006}%
  \BibitemOpen
  \bibfield  {author} {\bibinfo {author} {\bibfnamefont {M.}~\bibnamefont
  {Katsnelson}}, \bibinfo {author} {\bibfnamefont {K.}~\bibnamefont
  {Novoselov}},\ and\ \bibinfo {author} {\bibfnamefont {A.}~\bibnamefont
  {Geim}},\ }\bibfield  {title} {\bibinfo {title} {Chiral tunnelling and the
  klein paradox in graphene},\ }\href {https://doi.org/10.1038/nphys384}
  {\bibfield  {journal} {\bibinfo  {journal} {Nat. Phys.}\ }\textbf {\bibinfo
  {volume} {2}},\ \bibinfo {pages} {620} (\bibinfo {year} {2006})}\BibitemShut
  {NoStop}%
\bibitem [{\citenamefont {Novoselov}\ \emph {et~al.}(2005)\citenamefont
  {Novoselov}, \citenamefont {Geim}, \citenamefont {Morozov}, \citenamefont
  {Jiang}, \citenamefont {Katsnelson}, \citenamefont {Grigorieva},
  \citenamefont {Dubonos},\ and\ \citenamefont {Firsov}}]{Novoselov2005}%
  \BibitemOpen
  \bibfield  {author} {\bibinfo {author} {\bibfnamefont {K.~S.}\ \bibnamefont
  {Novoselov}}, \bibinfo {author} {\bibfnamefont {A.~K.}\ \bibnamefont {Geim}},
  \bibinfo {author} {\bibfnamefont {S.~V.}\ \bibnamefont {Morozov}}, \bibinfo
  {author} {\bibfnamefont {D.}~\bibnamefont {Jiang}}, \bibinfo {author}
  {\bibfnamefont {M.~I.}\ \bibnamefont {Katsnelson}}, \bibinfo {author}
  {\bibfnamefont {I.~V.}\ \bibnamefont {Grigorieva}}, \bibinfo {author}
  {\bibfnamefont {S.~V.}\ \bibnamefont {Dubonos}},\ and\ \bibinfo {author}
  {\bibfnamefont {A.~A.}\ \bibnamefont {Firsov}},\ }\bibfield  {title}
  {\bibinfo {title} {Two-dimensional gas of massless dirac fermions in
  graphene},\ }\href {https://doi.org/10.1038/nature04233} {\bibfield
  {journal} {\bibinfo  {journal} {Nature}\ }\textbf {\bibinfo {volume} {438}},\
  \bibinfo {pages} {197} (\bibinfo {year} {2005})}\BibitemShut {NoStop}%
\bibitem [{\citenamefont {Katsnelson}\ and\ \citenamefont
  {Novoselov}(2007)}]{Katsnelson2007}%
  \BibitemOpen
  \bibfield  {author} {\bibinfo {author} {\bibfnamefont {M.}~\bibnamefont
  {Katsnelson}}\ and\ \bibinfo {author} {\bibfnamefont {K.}~\bibnamefont
  {Novoselov}},\ }\bibfield  {title} {\bibinfo {title} {Graphene: New bridge
  between condensed matter physics and quantum electrodynamics},\ }\href
  {https://doi.org/10.1016/j.ssc.2007.02.043} {\bibfield  {journal} {\bibinfo
  {journal} {Solid State Commun.}\ }\textbf {\bibinfo {volume} {143}},\
  \bibinfo {pages} {3} (\bibinfo {year} {2007})}\BibitemShut {NoStop}%
\bibitem [{\citenamefont {McCann}\ and\ \citenamefont
  {Koshino}(2013)}]{McCann2013}%
  \BibitemOpen
  \bibfield  {author} {\bibinfo {author} {\bibfnamefont {E.}~\bibnamefont
  {McCann}}\ and\ \bibinfo {author} {\bibfnamefont {M.}~\bibnamefont
  {Koshino}},\ }\bibfield  {title} {\bibinfo {title} {The electronic properties
  of bilayer graphene},\ }\href {https://doi.org/10.1088/0034-4885/76/5/056503}
  {\bibfield  {journal} {\bibinfo  {journal} {Rep. Progr. Phys.}\ }\textbf
  {\bibinfo {volume} {76}},\ \bibinfo {pages} {056503} (\bibinfo {year}
  {2013})}\BibitemShut {NoStop}%
\bibitem [{\citenamefont {Kleptsyn}\ \emph {et~al.}(2015)\citenamefont
  {Kleptsyn}, \citenamefont {Okunev}, \citenamefont {Schurov}, \citenamefont
  {Zubov},\ and\ \citenamefont {Katsnelson}}]{Kleptsyn2015}%
  \BibitemOpen
  \bibfield  {author} {\bibinfo {author} {\bibfnamefont {V.}~\bibnamefont
  {Kleptsyn}}, \bibinfo {author} {\bibfnamefont {A.}~\bibnamefont {Okunev}},
  \bibinfo {author} {\bibfnamefont {I.}~\bibnamefont {Schurov}}, \bibinfo
  {author} {\bibfnamefont {D.}~\bibnamefont {Zubov}},\ and\ \bibinfo {author}
  {\bibfnamefont {M.~I.}\ \bibnamefont {Katsnelson}},\ }\bibfield  {title}
  {\bibinfo {title} {Chiral tunneling through generic one-dimensional potential
  barriers in bilayer graphene},\ }\href
  {https://doi.org/10.1103/physrevb.92.165407} {\bibfield  {journal} {\bibinfo
  {journal} {Phys. Rev. B}\ }\textbf {\bibinfo {volume} {92}},\ \bibinfo
  {pages} {165407} (\bibinfo {year} {2015})}\BibitemShut {NoStop}%
\bibitem [{\citenamefont {Du}\ \emph {et~al.}(2018)\citenamefont {Du},
  \citenamefont {Liu}, \citenamefont {Mohrmann}, \citenamefont {Wu},
  \citenamefont {Krupke}, \citenamefont {von Löhneysen}, \citenamefont
  {Richter},\ and\ \citenamefont {Danneau}}]{Du2018}%
  \BibitemOpen
  \bibfield  {author} {\bibinfo {author} {\bibfnamefont {R.}~\bibnamefont
  {Du}}, \bibinfo {author} {\bibfnamefont {M.-H.}\ \bibnamefont {Liu}},
  \bibinfo {author} {\bibfnamefont {J.}~\bibnamefont {Mohrmann}}, \bibinfo
  {author} {\bibfnamefont {F.}~\bibnamefont {Wu}}, \bibinfo {author}
  {\bibfnamefont {R.}~\bibnamefont {Krupke}}, \bibinfo {author} {\bibfnamefont
  {H.}~\bibnamefont {von Löhneysen}}, \bibinfo {author} {\bibfnamefont
  {K.}~\bibnamefont {Richter}},\ and\ \bibinfo {author} {\bibfnamefont
  {R.}~\bibnamefont {Danneau}},\ }\bibfield  {title} {\bibinfo {title} {Tuning
  anti-klein to klein tunneling in bilayer graphene},\ }\href
  {https://doi.org/10.1103/physrevlett.121.127706} {\bibfield  {journal}
  {\bibinfo  {journal} {Phys. Rev. Lett.}\ }\textbf {\bibinfo {volume} {121}},\
  \bibinfo {pages} {127706} (\bibinfo {year} {2018})}\BibitemShut {NoStop}%
\bibitem [{\citenamefont {Cao}\ \emph {et~al.}(2018{\natexlab{a}})\citenamefont
  {Cao}, \citenamefont {Fatemi}, \citenamefont {Fang}, \citenamefont
  {Watanabe}, \citenamefont {Taniguchi}, \citenamefont {Kaxiras},\ and\
  \citenamefont {Jarillo-Herrero}}]{Cao2018-1}%
  \BibitemOpen
  \bibfield  {author} {\bibinfo {author} {\bibfnamefont {Y.}~\bibnamefont
  {Cao}}, \bibinfo {author} {\bibfnamefont {V.}~\bibnamefont {Fatemi}},
  \bibinfo {author} {\bibfnamefont {S.}~\bibnamefont {Fang}}, \bibinfo {author}
  {\bibfnamefont {K.}~\bibnamefont {Watanabe}}, \bibinfo {author}
  {\bibfnamefont {T.}~\bibnamefont {Taniguchi}}, \bibinfo {author}
  {\bibfnamefont {E.}~\bibnamefont {Kaxiras}},\ and\ \bibinfo {author}
  {\bibfnamefont {P.}~\bibnamefont {Jarillo-Herrero}},\ }\bibfield  {title}
  {\bibinfo {title} {Unconventional superconductivity in magic-angle graphene
  superlattices},\ }\href {https://doi.org/10.1038/nature26160} {\bibfield
  {journal} {\bibinfo  {journal} {Nature}\ }\textbf {\bibinfo {volume} {556}},\
  \bibinfo {pages} {43} (\bibinfo {year} {2018}{\natexlab{a}})}\BibitemShut
  {NoStop}%
\bibitem [{\citenamefont {Cao}\ \emph {et~al.}(2018{\natexlab{b}})\citenamefont
  {Cao}, \citenamefont {Fatemi}, \citenamefont {Demir}, \citenamefont {Fang},
  \citenamefont {Tomarken}, \citenamefont {Luo}, \citenamefont
  {Sanchez-Yamagishi}, \citenamefont {Watanabe}, \citenamefont {Taniguchi},
  \citenamefont {Kaxiras} \emph {et~al.}}]{Cao2018-2}%
  \BibitemOpen
  \bibfield  {author} {\bibinfo {author} {\bibfnamefont {Y.}~\bibnamefont
  {Cao}}, \bibinfo {author} {\bibfnamefont {V.}~\bibnamefont {Fatemi}},
  \bibinfo {author} {\bibfnamefont {A.}~\bibnamefont {Demir}}, \bibinfo
  {author} {\bibfnamefont {S.}~\bibnamefont {Fang}}, \bibinfo {author}
  {\bibfnamefont {S.~L.}\ \bibnamefont {Tomarken}}, \bibinfo {author}
  {\bibfnamefont {J.~Y.}\ \bibnamefont {Luo}}, \bibinfo {author} {\bibfnamefont
  {J.~D.}\ \bibnamefont {Sanchez-Yamagishi}}, \bibinfo {author} {\bibfnamefont
  {K.}~\bibnamefont {Watanabe}}, \bibinfo {author} {\bibfnamefont
  {T.}~\bibnamefont {Taniguchi}}, \bibinfo {author} {\bibfnamefont
  {E.}~\bibnamefont {Kaxiras}}, \emph {et~al.},\ }\bibfield  {title} {\bibinfo
  {title} {Correlated insulator behaviour at half-filling in magic-angle
  graphene superlattices},\ }\href {https://doi.org/10.1038/nature26154}
  {\bibfield  {journal} {\bibinfo  {journal} {Nature}\ }\textbf {\bibinfo
  {volume} {556}},\ \bibinfo {pages} {80} (\bibinfo {year}
  {2018}{\natexlab{b}})}\BibitemShut {NoStop}%
\bibitem [{\citenamefont {MacDonald}(2019)}]{MacDonald2019}%
  \BibitemOpen
  \bibfield  {author} {\bibinfo {author} {\bibfnamefont {A.~H.}\ \bibnamefont
  {MacDonald}},\ }\bibfield  {title} {\bibinfo {title} {Bilayer graphene's
  wicked, twisted road},\ }\href {https://doi.org/10.1103/physics.12.12}
  {\bibfield  {journal} {\bibinfo  {journal} {Physics}\ }\textbf {\bibinfo
  {volume} {12}},\ \bibinfo {pages} {12} (\bibinfo {year} {2019})}\BibitemShut
  {NoStop}%
\bibitem [{\citenamefont {Catarina}\ \emph {et~al.}(2019)\citenamefont
  {Catarina}, \citenamefont {Amorim}, \citenamefont {Castro}, \citenamefont
  {Castro}, \citenamefont {Castro}, \citenamefont {Lopes}, \citenamefont
  {Lopes},\ and\ \citenamefont {Peres}}]{Catarina2019}%
  \BibitemOpen
  \bibfield  {author} {\bibinfo {author} {\bibfnamefont {G.}~\bibnamefont
  {Catarina}}, \bibinfo {author} {\bibfnamefont {B.}~\bibnamefont {Amorim}},
  \bibinfo {author} {\bibfnamefont {E.~V.}\ \bibnamefont {Castro}}, \bibinfo
  {author} {\bibfnamefont {E.~V.}\ \bibnamefont {Castro}}, \bibinfo {author}
  {\bibfnamefont {E.~V.}\ \bibnamefont {Castro}}, \bibinfo {author}
  {\bibfnamefont {J.~M. V.~P.}\ \bibnamefont {Lopes}}, \bibinfo {author}
  {\bibfnamefont {J.~M. V.~P.}\ \bibnamefont {Lopes}},\ and\ \bibinfo {author}
  {\bibfnamefont {N.}~\bibnamefont {Peres}},\ }\bibinfo {title} {Handbook of
  graphene}\ (\bibinfo  {publisher} {Wiley},\ \bibinfo {year} {2019})\ Chap.\
  \bibinfo {chapter} {Twisted Bilayer Graphene: Low-Energy Physics, Electronic
  and Optical Properties}, pp.\ \bibinfo {pages} {177--231}\BibitemShut
  {NoStop}%
\bibitem [{\citenamefont {Mogera}\ and\ \citenamefont
  {Kulkarni}(2020)}]{Mogera2020}%
  \BibitemOpen
  \bibfield  {author} {\bibinfo {author} {\bibfnamefont {U.}~\bibnamefont
  {Mogera}}\ and\ \bibinfo {author} {\bibfnamefont {G.~U.}\ \bibnamefont
  {Kulkarni}},\ }\bibfield  {title} {\bibinfo {title} {A new twist in graphene
  research: Twisted graphene},\ }\href
  {https://doi.org/10.1016/j.carbon.2019.09.053} {\bibfield  {journal}
  {\bibinfo  {journal} {Carbon}\ }\textbf {\bibinfo {volume} {156}},\ \bibinfo
  {pages} {470} (\bibinfo {year} {2020})}\BibitemShut {NoStop}%
\bibitem [{\citenamefont {Andrei}\ and\ \citenamefont
  {MacDonald}(2020)}]{Andrei2020}%
  \BibitemOpen
  \bibfield  {author} {\bibinfo {author} {\bibfnamefont {E.~Y.}\ \bibnamefont
  {Andrei}}\ and\ \bibinfo {author} {\bibfnamefont {A.~H.}\ \bibnamefont
  {MacDonald}},\ }\bibfield  {title} {\bibinfo {title} {Graphene bilayers with
  a twist},\ }\href {https://doi.org/10.1038/s41563-020-00840-0} {\bibfield
  {journal} {\bibinfo  {journal} {Nat. Mater.}\ }\textbf {\bibinfo {volume}
  {19}},\ \bibinfo {pages} {1265} (\bibinfo {year} {2020})}\BibitemShut
  {NoStop}%
\bibitem [{\citenamefont {Nimbalkar}\ and\ \citenamefont
  {Kim}(2020)}]{Nimbalkar2020}%
  \BibitemOpen
  \bibfield  {author} {\bibinfo {author} {\bibfnamefont {A.}~\bibnamefont
  {Nimbalkar}}\ and\ \bibinfo {author} {\bibfnamefont {H.}~\bibnamefont
  {Kim}},\ }\bibfield  {title} {\bibinfo {title} {Opportunities and challenges
  in twisted bilayer graphene: A review},\ }\href
  {https://doi.org/10.1007/s40820-020-00464-8} {\bibfield  {journal} {\bibinfo
  {journal} {Nano-Micro Lett.}\ }\textbf {\bibinfo {volume} {12}},\ \bibinfo
  {pages} {126} (\bibinfo {year} {2020})}\BibitemShut {NoStop}%
\bibitem [{\citenamefont {Carr}\ \emph {et~al.}(2017)\citenamefont {Carr},
  \citenamefont {Massatt}, \citenamefont {Fang}, \citenamefont {Cazeaux},
  \citenamefont {Luskin},\ and\ \citenamefont {Kaxiras}}]{Carr2017}%
  \BibitemOpen
  \bibfield  {author} {\bibinfo {author} {\bibfnamefont {S.}~\bibnamefont
  {Carr}}, \bibinfo {author} {\bibfnamefont {D.}~\bibnamefont {Massatt}},
  \bibinfo {author} {\bibfnamefont {S.}~\bibnamefont {Fang}}, \bibinfo {author}
  {\bibfnamefont {P.}~\bibnamefont {Cazeaux}}, \bibinfo {author} {\bibfnamefont
  {M.}~\bibnamefont {Luskin}},\ and\ \bibinfo {author} {\bibfnamefont
  {E.}~\bibnamefont {Kaxiras}},\ }\bibfield  {title} {\bibinfo {title}
  {Twistronics: Manipulating the electronic properties of two-dimensional
  layered structures through their twist angle},\ }\href
  {https://doi.org/10.1103/physrevb.95.075420} {\bibfield  {journal} {\bibinfo
  {journal} {Phys. Rev. B}\ }\textbf {\bibinfo {volume} {95}},\ \bibinfo
  {pages} {075420} (\bibinfo {year} {2017})}\BibitemShut {NoStop}%
\bibitem [{\citenamefont {Ribeiro-Palau}\ \emph {et~al.}(2018)\citenamefont
  {Ribeiro-Palau}, \citenamefont {Zhang}, \citenamefont {Watanabe},
  \citenamefont {Taniguchi}, \citenamefont {Hone},\ and\ \citenamefont
  {Dean}}]{Ribeiro-Palau2018}%
  \BibitemOpen
  \bibfield  {author} {\bibinfo {author} {\bibfnamefont {R.}~\bibnamefont
  {Ribeiro-Palau}}, \bibinfo {author} {\bibfnamefont {C.}~\bibnamefont
  {Zhang}}, \bibinfo {author} {\bibfnamefont {K.}~\bibnamefont {Watanabe}},
  \bibinfo {author} {\bibfnamefont {T.}~\bibnamefont {Taniguchi}}, \bibinfo
  {author} {\bibfnamefont {J.}~\bibnamefont {Hone}},\ and\ \bibinfo {author}
  {\bibfnamefont {C.~R.}\ \bibnamefont {Dean}},\ }\bibfield  {title} {\bibinfo
  {title} {Twistable electronics with dynamically rotatable heterostructures},\
  }\href {https://doi.org/10.1126/science.aat6981} {\bibfield  {journal}
  {\bibinfo  {journal} {Science}\ }\textbf {\bibinfo {volume} {361}},\ \bibinfo
  {pages} {690} (\bibinfo {year} {2018})}\BibitemShut {NoStop}%
\bibitem [{\citenamefont {Tarnopolsky}\ \emph {et~al.}(2019)\citenamefont
  {Tarnopolsky}, \citenamefont {Kruchkov},\ and\ \citenamefont
  {Vishwanath}}]{Tarnopolsky2019}%
  \BibitemOpen
  \bibfield  {author} {\bibinfo {author} {\bibfnamefont {G.}~\bibnamefont
  {Tarnopolsky}}, \bibinfo {author} {\bibfnamefont {A.~J.}\ \bibnamefont
  {Kruchkov}},\ and\ \bibinfo {author} {\bibfnamefont {A.}~\bibnamefont
  {Vishwanath}},\ }\bibfield  {title} {\bibinfo {title} {Origin of magic angles
  in twisted bilayer graphene},\ }\href
  {https://doi.org/10.1103/physrevlett.122.106405} {\bibfield  {journal}
  {\bibinfo  {journal} {Phys. Rev. Lett.}\ }\textbf {\bibinfo {volume} {122}},\
  \bibinfo {pages} {106405} (\bibinfo {year} {2019})}\BibitemShut {NoStop}%
\bibitem [{\citenamefont {Lisi}\ \emph {et~al.}(2020)\citenamefont {Lisi},
  \citenamefont {Lu}, \citenamefont {Benschop}, \citenamefont {de~Jong},
  \citenamefont {Stepanov}, \citenamefont {Duran}, \citenamefont {Margot},
  \citenamefont {Cucchi}, \citenamefont {Cappelli}, \citenamefont {Hunter},
  \citenamefont {Tamai}, \citenamefont {Kandyba}, \citenamefont {Giampietri},
  \citenamefont {Barinov}, \citenamefont {Jobst}, \citenamefont {Stalman},
  \citenamefont {Leeuwenhoek}, \citenamefont {Watanabe}, \citenamefont
  {Taniguchi}, \citenamefont {Rademaker}, \citenamefont {van~der Molen},
  \citenamefont {Allan}, \citenamefont {Efetov},\ and\ \citenamefont
  {Baumberger}}]{Lisi2020}%
  \BibitemOpen
  \bibfield  {author} {\bibinfo {author} {\bibfnamefont {S.}~\bibnamefont
  {Lisi}}, \bibinfo {author} {\bibfnamefont {X.}~\bibnamefont {Lu}}, \bibinfo
  {author} {\bibfnamefont {T.}~\bibnamefont {Benschop}}, \bibinfo {author}
  {\bibfnamefont {T.~A.}\ \bibnamefont {de~Jong}}, \bibinfo {author}
  {\bibfnamefont {P.}~\bibnamefont {Stepanov}}, \bibinfo {author}
  {\bibfnamefont {J.~R.}\ \bibnamefont {Duran}}, \bibinfo {author}
  {\bibfnamefont {F.}~\bibnamefont {Margot}}, \bibinfo {author} {\bibfnamefont
  {I.}~\bibnamefont {Cucchi}}, \bibinfo {author} {\bibfnamefont
  {E.}~\bibnamefont {Cappelli}}, \bibinfo {author} {\bibfnamefont
  {A.}~\bibnamefont {Hunter}}, \bibinfo {author} {\bibfnamefont
  {A.}~\bibnamefont {Tamai}}, \bibinfo {author} {\bibfnamefont
  {V.}~\bibnamefont {Kandyba}}, \bibinfo {author} {\bibfnamefont
  {A.}~\bibnamefont {Giampietri}}, \bibinfo {author} {\bibfnamefont
  {A.}~\bibnamefont {Barinov}}, \bibinfo {author} {\bibfnamefont
  {J.}~\bibnamefont {Jobst}}, \bibinfo {author} {\bibfnamefont
  {V.}~\bibnamefont {Stalman}}, \bibinfo {author} {\bibfnamefont
  {M.}~\bibnamefont {Leeuwenhoek}}, \bibinfo {author} {\bibfnamefont
  {K.}~\bibnamefont {Watanabe}}, \bibinfo {author} {\bibfnamefont
  {T.}~\bibnamefont {Taniguchi}}, \bibinfo {author} {\bibfnamefont
  {L.}~\bibnamefont {Rademaker}}, \bibinfo {author} {\bibfnamefont {S.~J.}\
  \bibnamefont {van~der Molen}}, \bibinfo {author} {\bibfnamefont {M.~P.}\
  \bibnamefont {Allan}}, \bibinfo {author} {\bibfnamefont {D.~K.}\ \bibnamefont
  {Efetov}},\ and\ \bibinfo {author} {\bibfnamefont {F.}~\bibnamefont
  {Baumberger}},\ }\bibfield  {title} {\bibinfo {title} {Observation of flat
  bands in twisted bilayer graphene},\ }\href
  {https://doi.org/10.1038/s41567-020-01041-x} {\bibfield  {journal} {\bibinfo
  {journal} {Nat. Phys.}\ }\textbf {\bibinfo {volume} {17}},\ \bibinfo {pages}
  {189} (\bibinfo {year} {2020})}\BibitemShut {NoStop}%
\bibitem [{\citenamefont {Jiang}\ \emph {et~al.}(2019)\citenamefont {Jiang},
  \citenamefont {Lai}, \citenamefont {Watanabe}, \citenamefont {Taniguchi},
  \citenamefont {Haule}, \citenamefont {Mao},\ and\ \citenamefont
  {Andrei}}]{Jiang2019}%
  \BibitemOpen
  \bibfield  {author} {\bibinfo {author} {\bibfnamefont {Y.}~\bibnamefont
  {Jiang}}, \bibinfo {author} {\bibfnamefont {X.}~\bibnamefont {Lai}}, \bibinfo
  {author} {\bibfnamefont {K.}~\bibnamefont {Watanabe}}, \bibinfo {author}
  {\bibfnamefont {T.}~\bibnamefont {Taniguchi}}, \bibinfo {author}
  {\bibfnamefont {K.}~\bibnamefont {Haule}}, \bibinfo {author} {\bibfnamefont
  {J.}~\bibnamefont {Mao}},\ and\ \bibinfo {author} {\bibfnamefont {E.~Y.}\
  \bibnamefont {Andrei}},\ }\bibfield  {title} {\bibinfo {title} {Charge order
  and broken rotational symmetry in magic-angle twisted bilayer graphene},\
  }\href {https://doi.org/10.1038/s41586-019-1460-4} {\bibfield  {journal}
  {\bibinfo  {journal} {Nature}\ }\textbf {\bibinfo {volume} {573}},\ \bibinfo
  {pages} {91} (\bibinfo {year} {2019})}\BibitemShut {NoStop}%
\bibitem [{\citenamefont {Choi}\ \emph {et~al.}(2019)\citenamefont {Choi},
  \citenamefont {Kemmer}, \citenamefont {Peng}, \citenamefont {Thomson},
  \citenamefont {Arora}, \citenamefont {Polski}, \citenamefont {Zhang},
  \citenamefont {Ren}, \citenamefont {Alicea}, \citenamefont {Refael},
  \citenamefont {von Oppen}, \citenamefont {Watanabe}, \citenamefont
  {Taniguchi},\ and\ \citenamefont {Nadj-Perge}}]{Choi2019}%
  \BibitemOpen
  \bibfield  {author} {\bibinfo {author} {\bibfnamefont {Y.}~\bibnamefont
  {Choi}}, \bibinfo {author} {\bibfnamefont {J.}~\bibnamefont {Kemmer}},
  \bibinfo {author} {\bibfnamefont {Y.}~\bibnamefont {Peng}}, \bibinfo {author}
  {\bibfnamefont {A.}~\bibnamefont {Thomson}}, \bibinfo {author} {\bibfnamefont
  {H.}~\bibnamefont {Arora}}, \bibinfo {author} {\bibfnamefont
  {R.}~\bibnamefont {Polski}}, \bibinfo {author} {\bibfnamefont
  {Y.}~\bibnamefont {Zhang}}, \bibinfo {author} {\bibfnamefont
  {H.}~\bibnamefont {Ren}}, \bibinfo {author} {\bibfnamefont {J.}~\bibnamefont
  {Alicea}}, \bibinfo {author} {\bibfnamefont {G.}~\bibnamefont {Refael}},
  \bibinfo {author} {\bibfnamefont {F.}~\bibnamefont {von Oppen}}, \bibinfo
  {author} {\bibfnamefont {K.}~\bibnamefont {Watanabe}}, \bibinfo {author}
  {\bibfnamefont {T.}~\bibnamefont {Taniguchi}},\ and\ \bibinfo {author}
  {\bibfnamefont {S.}~\bibnamefont {Nadj-Perge}},\ }\bibfield  {title}
  {\bibinfo {title} {Electronic correlations in twisted bilayer graphene near
  the magic angle},\ }\href {https://doi.org/10.1038/s41567-019-0606-5}
  {\bibfield  {journal} {\bibinfo  {journal} {Nat. Phys.}\ }\textbf {\bibinfo
  {volume} {15}},\ \bibinfo {pages} {1174} (\bibinfo {year}
  {2019})}\BibitemShut {NoStop}%
\bibitem [{\citenamefont {Xie}\ \emph {et~al.}(2019)\citenamefont {Xie},
  \citenamefont {Lian}, \citenamefont {Jäck}, \citenamefont {Liu},
  \citenamefont {Chiu}, \citenamefont {Watanabe}, \citenamefont {Taniguchi},
  \citenamefont {Bernevig},\ and\ \citenamefont {Yazdani}}]{Xie2019}%
  \BibitemOpen
  \bibfield  {author} {\bibinfo {author} {\bibfnamefont {Y.}~\bibnamefont
  {Xie}}, \bibinfo {author} {\bibfnamefont {B.}~\bibnamefont {Lian}}, \bibinfo
  {author} {\bibfnamefont {B.}~\bibnamefont {Jäck}}, \bibinfo {author}
  {\bibfnamefont {X.}~\bibnamefont {Liu}}, \bibinfo {author} {\bibfnamefont
  {C.-L.}\ \bibnamefont {Chiu}}, \bibinfo {author} {\bibfnamefont
  {K.}~\bibnamefont {Watanabe}}, \bibinfo {author} {\bibfnamefont
  {T.}~\bibnamefont {Taniguchi}}, \bibinfo {author} {\bibfnamefont {B.~A.}\
  \bibnamefont {Bernevig}},\ and\ \bibinfo {author} {\bibfnamefont
  {A.}~\bibnamefont {Yazdani}},\ }\bibfield  {title} {\bibinfo {title}
  {Spectroscopic signatures of many-body correlations in magic-angle twisted
  bilayer graphene},\ }\href {https://doi.org/10.1038/s41586-019-1422-x}
  {\bibfield  {journal} {\bibinfo  {journal} {Nature}\ }\textbf {\bibinfo
  {volume} {572}},\ \bibinfo {pages} {101} (\bibinfo {year}
  {2019})}\BibitemShut {NoStop}%
\bibitem [{\citenamefont {Wong}\ \emph {et~al.}(2020)\citenamefont {Wong},
  \citenamefont {Nuckolls}, \citenamefont {Oh}, \citenamefont {Lian},
  \citenamefont {Xie}, \citenamefont {Jeon}, \citenamefont {Watanabe},
  \citenamefont {Taniguchi}, \citenamefont {Bernevig},\ and\ \citenamefont
  {Yazdani}}]{Wong2020}%
  \BibitemOpen
  \bibfield  {author} {\bibinfo {author} {\bibfnamefont {D.}~\bibnamefont
  {Wong}}, \bibinfo {author} {\bibfnamefont {K.~P.}\ \bibnamefont {Nuckolls}},
  \bibinfo {author} {\bibfnamefont {M.}~\bibnamefont {Oh}}, \bibinfo {author}
  {\bibfnamefont {B.}~\bibnamefont {Lian}}, \bibinfo {author} {\bibfnamefont
  {Y.}~\bibnamefont {Xie}}, \bibinfo {author} {\bibfnamefont {S.}~\bibnamefont
  {Jeon}}, \bibinfo {author} {\bibfnamefont {K.}~\bibnamefont {Watanabe}},
  \bibinfo {author} {\bibfnamefont {T.}~\bibnamefont {Taniguchi}}, \bibinfo
  {author} {\bibfnamefont {B.~A.}\ \bibnamefont {Bernevig}},\ and\ \bibinfo
  {author} {\bibfnamefont {A.}~\bibnamefont {Yazdani}},\ }\bibfield  {title}
  {\bibinfo {title} {Cascade of electronic transitions in magic-angle twisted
  bilayer graphene},\ }\href {https://doi.org/10.1038/s41586-020-2339-0}
  {\bibfield  {journal} {\bibinfo  {journal} {Nature}\ }\textbf {\bibinfo
  {volume} {582}},\ \bibinfo {pages} {198} (\bibinfo {year}
  {2020})}\BibitemShut {NoStop}%
\bibitem [{\citenamefont {Xu}\ \emph {et~al.}(2021)\citenamefont {Xu},
  \citenamefont {Ezzi}, \citenamefont {Balakrishnan}, \citenamefont
  {Garcia-Ruiz}, \citenamefont {Tsim}, \citenamefont {Mullan}, \citenamefont
  {Barrier}, \citenamefont {Xin}, \citenamefont {Piot}, \citenamefont
  {Taniguchi}, \citenamefont {Watanabe}, \citenamefont {Carvalho},
  \citenamefont {Mishchenko}, \citenamefont {Geim}, \citenamefont {Fal'ko},
  \citenamefont {Adam}, \citenamefont {Neto}, \citenamefont {Novoselov},\ and\
  \citenamefont {Shi}}]{Xu2021}%
  \BibitemOpen
  \bibfield  {author} {\bibinfo {author} {\bibfnamefont {S.}~\bibnamefont
  {Xu}}, \bibinfo {author} {\bibfnamefont {M.~M.~A.}\ \bibnamefont {Ezzi}},
  \bibinfo {author} {\bibfnamefont {N.}~\bibnamefont {Balakrishnan}}, \bibinfo
  {author} {\bibfnamefont {A.}~\bibnamefont {Garcia-Ruiz}}, \bibinfo {author}
  {\bibfnamefont {B.}~\bibnamefont {Tsim}}, \bibinfo {author} {\bibfnamefont
  {C.}~\bibnamefont {Mullan}}, \bibinfo {author} {\bibfnamefont
  {J.}~\bibnamefont {Barrier}}, \bibinfo {author} {\bibfnamefont
  {N.}~\bibnamefont {Xin}}, \bibinfo {author} {\bibfnamefont {B.~A.}\
  \bibnamefont {Piot}}, \bibinfo {author} {\bibfnamefont {T.}~\bibnamefont
  {Taniguchi}}, \bibinfo {author} {\bibfnamefont {K.}~\bibnamefont {Watanabe}},
  \bibinfo {author} {\bibfnamefont {A.}~\bibnamefont {Carvalho}}, \bibinfo
  {author} {\bibfnamefont {A.}~\bibnamefont {Mishchenko}}, \bibinfo {author}
  {\bibfnamefont {A.~K.}\ \bibnamefont {Geim}}, \bibinfo {author}
  {\bibfnamefont {V.~I.}\ \bibnamefont {Fal'ko}}, \bibinfo {author}
  {\bibfnamefont {S.}~\bibnamefont {Adam}}, \bibinfo {author} {\bibfnamefont
  {A.~H.~C.}\ \bibnamefont {Neto}}, \bibinfo {author} {\bibfnamefont {K.~S.}\
  \bibnamefont {Novoselov}},\ and\ \bibinfo {author} {\bibfnamefont
  {Y.}~\bibnamefont {Shi}},\ }\bibfield  {title} {\bibinfo {title} {Tunable van
  hove singularities and correlated states in twisted
  monolayer{\textendash}bilayer graphene},\ }\href
  {https://doi.org/10.1038/s41567-021-01172-9} {\bibfield  {journal} {\bibinfo
  {journal} {Nat. Phys.}\ }\textbf {\bibinfo {volume} {17}},\ \bibinfo {pages}
  {619} (\bibinfo {year} {2021})}\BibitemShut {NoStop}%
\bibitem [{\citenamefont {Choi}\ \emph {et~al.}(2021)\citenamefont {Choi},
  \citenamefont {Kim}, \citenamefont {Lewandowski}, \citenamefont {Peng},
  \citenamefont {Thomson}, \citenamefont {Polski}, \citenamefont {Zhang},
  \citenamefont {Watanabe}, \citenamefont {Taniguchi}, \citenamefont {Alicea},\
  and\ \citenamefont {Nadj-Perge}}]{Choi2021}%
  \BibitemOpen
  \bibfield  {author} {\bibinfo {author} {\bibfnamefont {Y.}~\bibnamefont
  {Choi}}, \bibinfo {author} {\bibfnamefont {H.}~\bibnamefont {Kim}}, \bibinfo
  {author} {\bibfnamefont {C.}~\bibnamefont {Lewandowski}}, \bibinfo {author}
  {\bibfnamefont {Y.}~\bibnamefont {Peng}}, \bibinfo {author} {\bibfnamefont
  {A.}~\bibnamefont {Thomson}}, \bibinfo {author} {\bibfnamefont
  {R.}~\bibnamefont {Polski}}, \bibinfo {author} {\bibfnamefont
  {Y.}~\bibnamefont {Zhang}}, \bibinfo {author} {\bibfnamefont
  {K.}~\bibnamefont {Watanabe}}, \bibinfo {author} {\bibfnamefont
  {T.}~\bibnamefont {Taniguchi}}, \bibinfo {author} {\bibfnamefont
  {J.}~\bibnamefont {Alicea}},\ and\ \bibinfo {author} {\bibfnamefont
  {S.}~\bibnamefont {Nadj-Perge}},\ }\bibfield  {title} {\bibinfo {title}
  {Interaction-driven band flattening and correlated phases in twisted bilayer
  graphene}} (\bibinfo {year} {2021}),\ \bibinfo {note}
  {arXiv:2102.02209}\BibitemShut {NoStop}%
\bibitem [{\citenamefont {Gadelha}\ \emph {et~al.}(2021)\citenamefont
  {Gadelha}, \citenamefont {Ohlberg}, \citenamefont {Rabelo}, \citenamefont
  {Neto}, \citenamefont {Vasconcelos}, \citenamefont {Campos}, \citenamefont
  {Lemos}, \citenamefont {Ornelas}, \citenamefont {Miranda}, \citenamefont
  {Nadas} \emph {et~al.}}]{Gadelha2021}%
  \BibitemOpen
  \bibfield  {author} {\bibinfo {author} {\bibfnamefont {A.~C.}\ \bibnamefont
  {Gadelha}}, \bibinfo {author} {\bibfnamefont {D.~A.}\ \bibnamefont
  {Ohlberg}}, \bibinfo {author} {\bibfnamefont {C.}~\bibnamefont {Rabelo}},
  \bibinfo {author} {\bibfnamefont {E.~G.}\ \bibnamefont {Neto}}, \bibinfo
  {author} {\bibfnamefont {T.~L.}\ \bibnamefont {Vasconcelos}}, \bibinfo
  {author} {\bibfnamefont {J.~L.}\ \bibnamefont {Campos}}, \bibinfo {author}
  {\bibfnamefont {J.~S.}\ \bibnamefont {Lemos}}, \bibinfo {author}
  {\bibfnamefont {V.}~\bibnamefont {Ornelas}}, \bibinfo {author} {\bibfnamefont
  {D.}~\bibnamefont {Miranda}}, \bibinfo {author} {\bibfnamefont
  {R.}~\bibnamefont {Nadas}}, \emph {et~al.},\ }\bibfield  {title} {\bibinfo
  {title} {Localization of lattice dynamics in low-angle twisted bilayer
  graphene},\ }\href {https://doi.org/10.1038/s41586-021-03252-5} {\bibfield
  {journal} {\bibinfo  {journal} {Nature}\ }\textbf {\bibinfo {volume} {590}},\
  \bibinfo {pages} {405} (\bibinfo {year} {2021})}\BibitemShut {NoStop}%
\bibitem [{\citenamefont {Kennes}\ \emph {et~al.}(2021)\citenamefont {Kennes},
  \citenamefont {Claassen}, \citenamefont {Xian}, \citenamefont {Georges},
  \citenamefont {Millis}, \citenamefont {Hone}, \citenamefont {Dean},
  \citenamefont {Basov}, \citenamefont {Pasupathy},\ and\ \citenamefont
  {Rubio}}]{Kennes2021}%
  \BibitemOpen
  \bibfield  {author} {\bibinfo {author} {\bibfnamefont {D.~M.}\ \bibnamefont
  {Kennes}}, \bibinfo {author} {\bibfnamefont {M.}~\bibnamefont {Claassen}},
  \bibinfo {author} {\bibfnamefont {L.}~\bibnamefont {Xian}}, \bibinfo {author}
  {\bibfnamefont {A.}~\bibnamefont {Georges}}, \bibinfo {author} {\bibfnamefont
  {A.~J.}\ \bibnamefont {Millis}}, \bibinfo {author} {\bibfnamefont
  {J.}~\bibnamefont {Hone}}, \bibinfo {author} {\bibfnamefont {C.~R.}\
  \bibnamefont {Dean}}, \bibinfo {author} {\bibfnamefont {D.~N.}\ \bibnamefont
  {Basov}}, \bibinfo {author} {\bibfnamefont {A.~N.}\ \bibnamefont
  {Pasupathy}},\ and\ \bibinfo {author} {\bibfnamefont {A.}~\bibnamefont
  {Rubio}},\ }\bibfield  {title} {\bibinfo {title} {Moir{\'{e}}
  heterostructures as a condensed-matter quantum simulator},\ }\href
  {https://doi.org/10.1038/s41567-020-01154-3} {\bibfield  {journal} {\bibinfo
  {journal} {Nat. Phys.}\ }\textbf {\bibinfo {volume} {17}},\ \bibinfo {pages}
  {155} (\bibinfo {year} {2021})}\BibitemShut {NoStop}%
\bibitem [{\citenamefont {Lu}\ \emph {et~al.}(2019)\citenamefont {Lu},
  \citenamefont {Stepanov}, \citenamefont {Yang}, \citenamefont {Xie},
  \citenamefont {Aamir}, \citenamefont {Das}, \citenamefont {Urgell},
  \citenamefont {Watanabe}, \citenamefont {Taniguchi}, \citenamefont {Zhang},
  \citenamefont {Bachtold}, \citenamefont {MacDonald},\ and\ \citenamefont
  {Efetov}}]{Lu2019}%
  \BibitemOpen
  \bibfield  {author} {\bibinfo {author} {\bibfnamefont {X.}~\bibnamefont
  {Lu}}, \bibinfo {author} {\bibfnamefont {P.}~\bibnamefont {Stepanov}},
  \bibinfo {author} {\bibfnamefont {W.}~\bibnamefont {Yang}}, \bibinfo {author}
  {\bibfnamefont {M.}~\bibnamefont {Xie}}, \bibinfo {author} {\bibfnamefont
  {M.~A.}\ \bibnamefont {Aamir}}, \bibinfo {author} {\bibfnamefont
  {I.}~\bibnamefont {Das}}, \bibinfo {author} {\bibfnamefont {C.}~\bibnamefont
  {Urgell}}, \bibinfo {author} {\bibfnamefont {K.}~\bibnamefont {Watanabe}},
  \bibinfo {author} {\bibfnamefont {T.}~\bibnamefont {Taniguchi}}, \bibinfo
  {author} {\bibfnamefont {G.}~\bibnamefont {Zhang}}, \bibinfo {author}
  {\bibfnamefont {A.}~\bibnamefont {Bachtold}}, \bibinfo {author}
  {\bibfnamefont {A.~H.}\ \bibnamefont {MacDonald}},\ and\ \bibinfo {author}
  {\bibfnamefont {D.~K.}\ \bibnamefont {Efetov}},\ }\bibfield  {title}
  {\bibinfo {title} {Superconductors, orbital magnets and correlated states in
  magic-angle bilayer graphene},\ }\href
  {https://doi.org/10.1038/s41586-019-1695-0} {\bibfield  {journal} {\bibinfo
  {journal} {Nature}\ }\textbf {\bibinfo {volume} {574}},\ \bibinfo {pages}
  {653} (\bibinfo {year} {2019})}\BibitemShut {NoStop}%
\bibitem [{\citenamefont {Sharpe}\ \emph {et~al.}(2019)\citenamefont {Sharpe},
  \citenamefont {Fox}, \citenamefont {Barnard}, \citenamefont {Finney},
  \citenamefont {Watanabe}, \citenamefont {Taniguchi}, \citenamefont
  {Kastner},\ and\ \citenamefont {Goldhaber-Gordon}}]{Sharpe2019}%
  \BibitemOpen
  \bibfield  {author} {\bibinfo {author} {\bibfnamefont {A.~L.}\ \bibnamefont
  {Sharpe}}, \bibinfo {author} {\bibfnamefont {E.~J.}\ \bibnamefont {Fox}},
  \bibinfo {author} {\bibfnamefont {A.~W.}\ \bibnamefont {Barnard}}, \bibinfo
  {author} {\bibfnamefont {J.}~\bibnamefont {Finney}}, \bibinfo {author}
  {\bibfnamefont {K.}~\bibnamefont {Watanabe}}, \bibinfo {author}
  {\bibfnamefont {T.}~\bibnamefont {Taniguchi}}, \bibinfo {author}
  {\bibfnamefont {M.~A.}\ \bibnamefont {Kastner}},\ and\ \bibinfo {author}
  {\bibfnamefont {D.}~\bibnamefont {Goldhaber-Gordon}},\ }\bibfield  {title}
  {\bibinfo {title} {Emergent ferromagnetism near three-quarters filling in
  twisted bilayer graphene},\ }\href {https://doi.org/10.1126/science.aaw3780}
  {\bibfield  {journal} {\bibinfo  {journal} {Science}\ }\textbf {\bibinfo
  {volume} {365}},\ \bibinfo {pages} {605} (\bibinfo {year}
  {2019})}\BibitemShut {NoStop}%
\bibitem [{\citenamefont {Song}\ \emph {et~al.}(2019)\citenamefont {Song},
  \citenamefont {Wang}, \citenamefont {Shi}, \citenamefont {Li}, \citenamefont
  {Fang},\ and\ \citenamefont {Bernevig}}]{Song2019}%
  \BibitemOpen
  \bibfield  {author} {\bibinfo {author} {\bibfnamefont {Z.}~\bibnamefont
  {Song}}, \bibinfo {author} {\bibfnamefont {Z.}~\bibnamefont {Wang}}, \bibinfo
  {author} {\bibfnamefont {W.}~\bibnamefont {Shi}}, \bibinfo {author}
  {\bibfnamefont {G.}~\bibnamefont {Li}}, \bibinfo {author} {\bibfnamefont
  {C.}~\bibnamefont {Fang}},\ and\ \bibinfo {author} {\bibfnamefont {B.~A.}\
  \bibnamefont {Bernevig}},\ }\bibfield  {title} {\bibinfo {title} {All magic
  angles in twisted bilayer graphene are topological},\ }\href
  {https://doi.org/10.1103/physrevlett.123.036401} {\bibfield  {journal}
  {\bibinfo  {journal} {Phys. Rev. Lett.}\ }\textbf {\bibinfo {volume} {123}},\
  \bibinfo {pages} {036401} (\bibinfo {year} {2019})}\BibitemShut {NoStop}%
\bibitem [{\citenamefont {Nuckolls}\ \emph {et~al.}(2020)\citenamefont
  {Nuckolls}, \citenamefont {Oh}, \citenamefont {Wong}, \citenamefont {Lian},
  \citenamefont {Watanabe}, \citenamefont {Taniguchi}, \citenamefont
  {Bernevig},\ and\ \citenamefont {Yazdani}}]{Nuckolls2020}%
  \BibitemOpen
  \bibfield  {author} {\bibinfo {author} {\bibfnamefont {K.~P.}\ \bibnamefont
  {Nuckolls}}, \bibinfo {author} {\bibfnamefont {M.}~\bibnamefont {Oh}},
  \bibinfo {author} {\bibfnamefont {D.}~\bibnamefont {Wong}}, \bibinfo {author}
  {\bibfnamefont {B.}~\bibnamefont {Lian}}, \bibinfo {author} {\bibfnamefont
  {K.}~\bibnamefont {Watanabe}}, \bibinfo {author} {\bibfnamefont
  {T.}~\bibnamefont {Taniguchi}}, \bibinfo {author} {\bibfnamefont {B.~A.}\
  \bibnamefont {Bernevig}},\ and\ \bibinfo {author} {\bibfnamefont
  {A.}~\bibnamefont {Yazdani}},\ }\bibfield  {title} {\bibinfo {title}
  {Strongly correlated chern insulators in magic-angle twisted bilayer
  graphene},\ }\href {https://doi.org/10.1038/s41586-020-3028-8} {\bibfield
  {journal} {\bibinfo  {journal} {Nature}\ }\textbf {\bibinfo {volume} {588}},\
  \bibinfo {pages} {610} (\bibinfo {year} {2020})}\BibitemShut {NoStop}%
\bibitem [{\citenamefont {Bultinck}\ \emph {et~al.}(2020)\citenamefont
  {Bultinck}, \citenamefont {Chatterjee},\ and\ \citenamefont
  {Zaletel}}]{Bultinck2020}%
  \BibitemOpen
  \bibfield  {author} {\bibinfo {author} {\bibfnamefont {N.}~\bibnamefont
  {Bultinck}}, \bibinfo {author} {\bibfnamefont {S.}~\bibnamefont
  {Chatterjee}},\ and\ \bibinfo {author} {\bibfnamefont {M.~P.}\ \bibnamefont
  {Zaletel}},\ }\bibfield  {title} {\bibinfo {title} {Mechanism for anomalous
  hall ferromagnetism in twisted bilayer graphene},\ }\href
  {https://doi.org/10.1103/physrevlett.124.166601} {\bibfield  {journal}
  {\bibinfo  {journal} {Phys. Rev. Lett.}\ }\textbf {\bibinfo {volume} {124}},\
  \bibinfo {pages} {166601} (\bibinfo {year} {2020})}\BibitemShut {NoStop}%
\bibitem [{\citenamefont {Ma}\ \emph {et~al.}(2020)\citenamefont {Ma},
  \citenamefont {Wang}, \citenamefont {Mills}, \citenamefont {Chen},
  \citenamefont {Deng}, \citenamefont {Yuan}, \citenamefont {Li}, \citenamefont
  {Watanabe}, \citenamefont {Taniguchi}, \citenamefont {Du}, \citenamefont
  {Zhang},\ and\ \citenamefont {Xia}}]{Ma2020}%
  \BibitemOpen
  \bibfield  {author} {\bibinfo {author} {\bibfnamefont {C.}~\bibnamefont
  {Ma}}, \bibinfo {author} {\bibfnamefont {Q.}~\bibnamefont {Wang}}, \bibinfo
  {author} {\bibfnamefont {S.}~\bibnamefont {Mills}}, \bibinfo {author}
  {\bibfnamefont {X.}~\bibnamefont {Chen}}, \bibinfo {author} {\bibfnamefont
  {B.}~\bibnamefont {Deng}}, \bibinfo {author} {\bibfnamefont {S.}~\bibnamefont
  {Yuan}}, \bibinfo {author} {\bibfnamefont {C.}~\bibnamefont {Li}}, \bibinfo
  {author} {\bibfnamefont {K.}~\bibnamefont {Watanabe}}, \bibinfo {author}
  {\bibfnamefont {T.}~\bibnamefont {Taniguchi}}, \bibinfo {author}
  {\bibfnamefont {X.}~\bibnamefont {Du}}, \bibinfo {author} {\bibfnamefont
  {F.}~\bibnamefont {Zhang}},\ and\ \bibinfo {author} {\bibfnamefont
  {F.}~\bibnamefont {Xia}},\ }\bibfield  {title} {\bibinfo {title} {Moir{\'{e}}
  band topology in twisted bilayer graphene},\ }\href
  {https://doi.org/10.1021/acs.nanolett.0c02131} {\bibfield  {journal}
  {\bibinfo  {journal} {Nano Lett.}\ }\textbf {\bibinfo {volume} {20}},\
  \bibinfo {pages} {6076} (\bibinfo {year} {2020})}\BibitemShut {NoStop}%
\bibitem [{\citenamefont {Dai}\ \emph {et~al.}(2016)\citenamefont {Dai},
  \citenamefont {Xiang},\ and\ \citenamefont {Srolovitz}}]{Dai2016}%
  \BibitemOpen
  \bibfield  {author} {\bibinfo {author} {\bibfnamefont {S.}~\bibnamefont
  {Dai}}, \bibinfo {author} {\bibfnamefont {Y.}~\bibnamefont {Xiang}},\ and\
  \bibinfo {author} {\bibfnamefont {D.~J.}\ \bibnamefont {Srolovitz}},\
  }\bibfield  {title} {\bibinfo {title} {Twisted bilayer graphene: Moir{\'{e}}
  with a twist},\ }\href {https://doi.org/10.1021/acs.nanolett.6b02870}
  {\bibfield  {journal} {\bibinfo  {journal} {Nano Lett.}\ }\textbf {\bibinfo
  {volume} {16}},\ \bibinfo {pages} {5923} (\bibinfo {year}
  {2016})}\BibitemShut {NoStop}%
\bibitem [{\citenamefont {Walet}\ and\ \citenamefont
  {Guinea}(2019)}]{Walet2019}%
  \BibitemOpen
  \bibfield  {author} {\bibinfo {author} {\bibfnamefont {N.~R.}\ \bibnamefont
  {Walet}}\ and\ \bibinfo {author} {\bibfnamefont {F.}~\bibnamefont {Guinea}},\
  }\bibfield  {title} {\bibinfo {title} {The emergence of one-dimensional
  channels in marginal-angle twisted bilayer graphene},\ }\href
  {https://doi.org/10.1088/2053-1583/ab57f8} {\bibfield  {journal} {\bibinfo
  {journal} {2D Mater.}\ }\textbf {\bibinfo {volume} {7}},\ \bibinfo {pages}
  {015023} (\bibinfo {year} {2019})}\BibitemShut {NoStop}%
\bibitem [{\citenamefont {Fleischmann}\ \emph {et~al.}(2019)\citenamefont
  {Fleischmann}, \citenamefont {Gupta}, \citenamefont {Wullschläger},
  \citenamefont {Theil}, \citenamefont {Weckbecker}, \citenamefont {Meded},
  \citenamefont {Sharma}, \citenamefont {Meyer},\ and\ \citenamefont
  {Shallcross}}]{Fleischmann2019}%
  \BibitemOpen
  \bibfield  {author} {\bibinfo {author} {\bibfnamefont {M.}~\bibnamefont
  {Fleischmann}}, \bibinfo {author} {\bibfnamefont {R.}~\bibnamefont {Gupta}},
  \bibinfo {author} {\bibfnamefont {F.}~\bibnamefont {Wullschläger}}, \bibinfo
  {author} {\bibfnamefont {S.}~\bibnamefont {Theil}}, \bibinfo {author}
  {\bibfnamefont {D.}~\bibnamefont {Weckbecker}}, \bibinfo {author}
  {\bibfnamefont {V.}~\bibnamefont {Meded}}, \bibinfo {author} {\bibfnamefont
  {S.}~\bibnamefont {Sharma}}, \bibinfo {author} {\bibfnamefont
  {B.}~\bibnamefont {Meyer}},\ and\ \bibinfo {author} {\bibfnamefont
  {S.}~\bibnamefont {Shallcross}},\ }\bibfield  {title} {\bibinfo {title}
  {Perfect and controllable nesting in minimally twisted bilayer graphene},\
  }\href {https://doi.org/10.1021/acs.nanolett.9b04027} {\bibfield  {journal}
  {\bibinfo  {journal} {Nano Lett.}\ }\textbf {\bibinfo {volume} {20}},\
  \bibinfo {pages} {971} (\bibinfo {year} {2019})}\BibitemShut {NoStop}%
\bibitem [{\citenamefont {Nguyen}\ \emph {et~al.}(2021)\citenamefont {Nguyen},
  \citenamefont {Paszko}, \citenamefont {Lamparski}, \citenamefont {Troeye},
  \citenamefont {Meunier},\ and\ \citenamefont {Charlier}}]{Nguyen2021}%
  \BibitemOpen
  \bibfield  {author} {\bibinfo {author} {\bibfnamefont {V.~H.}\ \bibnamefont
  {Nguyen}}, \bibinfo {author} {\bibfnamefont {D.}~\bibnamefont {Paszko}},
  \bibinfo {author} {\bibfnamefont {M.}~\bibnamefont {Lamparski}}, \bibinfo
  {author} {\bibfnamefont {B.~V.}\ \bibnamefont {Troeye}}, \bibinfo {author}
  {\bibfnamefont {V.}~\bibnamefont {Meunier}},\ and\ \bibinfo {author}
  {\bibfnamefont {J.-C.}\ \bibnamefont {Charlier}},\ }\bibfield  {title}
  {\bibinfo {title} {Electronic localization in small-angle twisted bilayer
  graphene},\ }\href {https://doi.org/10.1088/2053-1583/ac044f} {\bibfield
  {journal} {\bibinfo  {journal} {2D Materials}\ }\textbf {\bibinfo {volume}
  {8}},\ \bibinfo {pages} {035046} (\bibinfo {year} {2021})}\BibitemShut
  {NoStop}%
\bibitem [{\citenamefont {Kim}\ \emph {et~al.}(2016)\citenamefont {Kim},
  \citenamefont {Herlinger}, \citenamefont {Moon}, \citenamefont {Koshino},
  \citenamefont {Taniguchi}, \citenamefont {Watanabe},\ and\ \citenamefont
  {Smet}}]{Kim2016}%
  \BibitemOpen
  \bibfield  {author} {\bibinfo {author} {\bibfnamefont {Y.}~\bibnamefont
  {Kim}}, \bibinfo {author} {\bibfnamefont {P.}~\bibnamefont {Herlinger}},
  \bibinfo {author} {\bibfnamefont {P.}~\bibnamefont {Moon}}, \bibinfo {author}
  {\bibfnamefont {M.}~\bibnamefont {Koshino}}, \bibinfo {author} {\bibfnamefont
  {T.}~\bibnamefont {Taniguchi}}, \bibinfo {author} {\bibfnamefont
  {K.}~\bibnamefont {Watanabe}},\ and\ \bibinfo {author} {\bibfnamefont
  {J.~H.}\ \bibnamefont {Smet}},\ }\bibfield  {title} {\bibinfo {title} {Charge
  inversion and topological phase transition at a twist angle induced van hove
  singularity of bilayer graphene},\ }\href
  {https://doi.org/10.1021/acs.nanolett.6b01906} {\bibfield  {journal}
  {\bibinfo  {journal} {Nano Lett.}\ }\textbf {\bibinfo {volume} {16}},\
  \bibinfo {pages} {5053} (\bibinfo {year} {2016})}\BibitemShut {NoStop}%
\bibitem [{\citenamefont {Hou}\ \emph {et~al.}(2020)\citenamefont {Hou},
  \citenamefont {Ren}, \citenamefont {Quan}, \citenamefont {Jung},
  \citenamefont {Ren},\ and\ \citenamefont {Qiao}}]{Hou2020}%
  \BibitemOpen
  \bibfield  {author} {\bibinfo {author} {\bibfnamefont {T.}~\bibnamefont
  {Hou}}, \bibinfo {author} {\bibfnamefont {Y.}~\bibnamefont {Ren}}, \bibinfo
  {author} {\bibfnamefont {Y.}~\bibnamefont {Quan}}, \bibinfo {author}
  {\bibfnamefont {J.}~\bibnamefont {Jung}}, \bibinfo {author} {\bibfnamefont
  {W.}~\bibnamefont {Ren}},\ and\ \bibinfo {author} {\bibfnamefont
  {Z.}~\bibnamefont {Qiao}},\ }\bibfield  {title} {\bibinfo {title} {Valley
  current splitter in minimally twisted bilayer graphene},\ }\href
  {https://doi.org/10.1103/physrevb.102.085433} {\bibfield  {journal} {\bibinfo
   {journal} {Phys. Rev. B}\ }\textbf {\bibinfo {volume} {102}},\ \bibinfo
  {pages} {085433} (\bibinfo {year} {2020})}\BibitemShut {NoStop}%
\bibitem [{\citenamefont {Schaibley}\ \emph {et~al.}(2016)\citenamefont
  {Schaibley}, \citenamefont {Yu}, \citenamefont {Clark}, \citenamefont
  {Rivera}, \citenamefont {Ross}, \citenamefont {Seyler}, \citenamefont {Yao},\
  and\ \citenamefont {Xu}}]{Schaibley2016}%
  \BibitemOpen
  \bibfield  {author} {\bibinfo {author} {\bibfnamefont {J.~R.}\ \bibnamefont
  {Schaibley}}, \bibinfo {author} {\bibfnamefont {H.}~\bibnamefont {Yu}},
  \bibinfo {author} {\bibfnamefont {G.}~\bibnamefont {Clark}}, \bibinfo
  {author} {\bibfnamefont {P.}~\bibnamefont {Rivera}}, \bibinfo {author}
  {\bibfnamefont {J.~S.}\ \bibnamefont {Ross}}, \bibinfo {author}
  {\bibfnamefont {K.~L.}\ \bibnamefont {Seyler}}, \bibinfo {author}
  {\bibfnamefont {W.}~\bibnamefont {Yao}},\ and\ \bibinfo {author}
  {\bibfnamefont {X.}~\bibnamefont {Xu}},\ }\bibfield  {title} {\bibinfo
  {title} {Valleytronics in 2d materials},\ }\href
  {https://doi.org/10.1038/natrevmats.2016.55} {\bibfield  {journal} {\bibinfo
  {journal} {Nat. Rev. Mater.}\ }\textbf {\bibinfo {volume} {1}},\ \bibinfo
  {pages} {16055} (\bibinfo {year} {2016})}\BibitemShut {NoStop}%
\bibitem [{\citenamefont {Han}\ \emph {et~al.}(2020)\citenamefont {Han},
  \citenamefont {Zeng}, \citenamefont {Ren}, \citenamefont {Dong},
  \citenamefont {Ren},\ and\ \citenamefont {Qiao}}]{Han2020}%
  \BibitemOpen
  \bibfield  {author} {\bibinfo {author} {\bibfnamefont {Y.}~\bibnamefont
  {Han}}, \bibinfo {author} {\bibfnamefont {J.}~\bibnamefont {Zeng}}, \bibinfo
  {author} {\bibfnamefont {Y.}~\bibnamefont {Ren}}, \bibinfo {author}
  {\bibfnamefont {X.}~\bibnamefont {Dong}}, \bibinfo {author} {\bibfnamefont
  {W.}~\bibnamefont {Ren}},\ and\ \bibinfo {author} {\bibfnamefont
  {Z.}~\bibnamefont {Qiao}},\ }\bibfield  {title} {\bibinfo {title} {Mesoscopic
  electronic transport in twisted bilayer graphene},\ }\href
  {https://doi.org/10.1103/physrevb.101.235432} {\bibfield  {journal} {\bibinfo
   {journal} {Phys. Rev. B}\ }\textbf {\bibinfo {volume} {101}},\ \bibinfo
  {pages} {235432} (\bibinfo {year} {2020})}\BibitemShut {NoStop}%
\bibitem [{\citenamefont {Datta}(1997)}]{Datta1997}%
  \BibitemOpen
  \bibfield  {author} {\bibinfo {author} {\bibfnamefont {S.}~\bibnamefont
  {Datta}},\ }\href@noop {} {\emph {\bibinfo {title} {Electronic Transport in
  Mesoscopic Systems}}}\ (\bibinfo  {publisher} {Cambridge University Press},\
  \bibinfo {year} {1997})\BibitemShut {NoStop}%
\bibitem [{\citenamefont {Datta}(2005)}]{Datta2005}%
  \BibitemOpen
  \bibfield  {author} {\bibinfo {author} {\bibfnamefont {S.}~\bibnamefont
  {Datta}},\ }\href@noop {} {\emph {\bibinfo {title} {Quantum Transport: Atom
  to Transistor}}}\ (\bibinfo  {publisher} {Cambridge University Press},\
  \bibinfo {year} {2005})\BibitemShut {NoStop}%
\bibitem [{\citenamefont {Di~Ventra}(2008)}]{DiVentra2008}%
  \BibitemOpen
  \bibfield  {author} {\bibinfo {author} {\bibfnamefont {M.}~\bibnamefont
  {Di~Ventra}},\ }\href@noop {} {\emph {\bibinfo {title} {Electrical Transport
  in Nanoscale Systems}}},\ \bibinfo {edition} {1st}\ ed.\ (\bibinfo
  {publisher} {Cambridge University Press},\ \bibinfo {year}
  {2008})\BibitemShut {NoStop}%
\bibitem [{\citenamefont {Verzijl}\ \emph {et~al.}(2013)\citenamefont
  {Verzijl}, \citenamefont {Seldenthuis},\ and\ \citenamefont
  {Thijssen}}]{Verzijl2013}%
  \BibitemOpen
  \bibfield  {author} {\bibinfo {author} {\bibfnamefont {C.~J.~O.}\
  \bibnamefont {Verzijl}}, \bibinfo {author} {\bibfnamefont {J.~S.}\
  \bibnamefont {Seldenthuis}},\ and\ \bibinfo {author} {\bibfnamefont {J.~M.}\
  \bibnamefont {Thijssen}},\ }\bibfield  {title} {\bibinfo {title}
  {Applicability of the wide-band limit in dft-based molecular transport
  calculations},\ }\href {https://doi.org/http://dx.doi.org/10.1063/1.4793259}
  {\bibfield  {journal} {\bibinfo  {journal} {J. Chem. Phys.}\ }\textbf
  {\bibinfo {volume} {138}},\ \bibinfo {eid} {094102} (\bibinfo {year}
  {2013})}\BibitemShut {NoStop}%
\bibitem [{\citenamefont {Stegmann}(2014)}]{StegmannPhd2014}%
  \BibitemOpen
  \bibfield  {author} {\bibinfo {author} {\bibfnamefont {T.}~\bibnamefont
  {Stegmann}},\ }\emph {\bibinfo {title} {Quantum transport in nanostructures:
  From the effects of decoherence on localization to magnetotransport in
  two-dimensional electron systems}},\ \href
  {http://duepublico.uni-duisburg-essen.de/servlets/DocumentServlet?id=35708&lang=en}
  {Ph.D. thesis},\ \bibinfo  {school} {University Duisburg-Essen} (\bibinfo
  {year} {2014})\BibitemShut {NoStop}%
\bibitem [{\citenamefont {Stegmann}\ and\ \citenamefont
  {Lorke}(2015)}]{Stegmann2015}%
  \BibitemOpen
  \bibfield  {author} {\bibinfo {author} {\bibfnamefont {T.}~\bibnamefont
  {Stegmann}}\ and\ \bibinfo {author} {\bibfnamefont {A.}~\bibnamefont
  {Lorke}},\ }\bibfield  {title} {\bibinfo {title} {Edge magnetotransport in
  graphene: A combined analytical and numerical study},\ }\href
  {https://doi.org/10.1002/andp.201500124} {\bibfield  {journal} {\bibinfo
  {journal} {Ann. Phys.}\ }\textbf {\bibinfo {volume} {527}},\ \bibinfo {pages}
  {723} (\bibinfo {year} {2015})}\BibitemShut {NoStop}%
\bibitem [{\citenamefont {Stegmann}\ and\ \citenamefont
  {Szpak}(2019)}]{Stegmann2019}%
  \BibitemOpen
  \bibfield  {author} {\bibinfo {author} {\bibfnamefont {T.}~\bibnamefont
  {Stegmann}}\ and\ \bibinfo {author} {\bibfnamefont {N.}~\bibnamefont
  {Szpak}},\ }\bibfield  {title} {\bibinfo {title} {Current splitting and
  valley polarization in elastically deformed graphene},\ }\href
  {https://doi.org/10.1088/2053-1583/aaea8d} {\bibfield  {journal} {\bibinfo
  {journal} {2D Mater.}\ }\textbf {\bibinfo {volume} {6}},\ \bibinfo {pages}
  {015024} (\bibinfo {year} {2019})}\BibitemShut {NoStop}%
\bibitem [{\citenamefont {Settnes}\ \emph {et~al.}(2016)\citenamefont
  {Settnes}, \citenamefont {Power}, \citenamefont {Brandbyge},\ and\
  \citenamefont {Jauho}}]{Settnes2016}%
  \BibitemOpen
  \bibfield  {author} {\bibinfo {author} {\bibfnamefont {M.}~\bibnamefont
  {Settnes}}, \bibinfo {author} {\bibfnamefont {S.~R.}\ \bibnamefont {Power}},
  \bibinfo {author} {\bibfnamefont {M.}~\bibnamefont {Brandbyge}},\ and\
  \bibinfo {author} {\bibfnamefont {A.-P.}\ \bibnamefont {Jauho}},\ }\bibfield
  {title} {\bibinfo {title} {Graphene nanobubbles as valley filters and beam
  splitters},\ }\href {https://doi.org/10.1103/PhysRevLett.117.276801}
  {\bibfield  {journal} {\bibinfo  {journal} {Phys. Rev. Lett.}\ }\textbf
  {\bibinfo {volume} {117}},\ \bibinfo {pages} {276801} (\bibinfo {year}
  {2016})}\BibitemShut {NoStop}%
\bibitem [{\citenamefont {Garcia-Pomar}\ \emph {et~al.}(2008)\citenamefont
  {Garcia-Pomar}, \citenamefont {Cortijo},\ and\ \citenamefont
  {Nieto-Vesperinas}}]{Garcia-Pomar2008}%
  \BibitemOpen
  \bibfield  {author} {\bibinfo {author} {\bibfnamefont {J.~L.}\ \bibnamefont
  {Garcia-Pomar}}, \bibinfo {author} {\bibfnamefont {A.}~\bibnamefont
  {Cortijo}},\ and\ \bibinfo {author} {\bibfnamefont {M.}~\bibnamefont
  {Nieto-Vesperinas}},\ }\bibfield  {title} {\bibinfo {title} {Fully
  valley-polarized electron beams in graphene},\ }\href
  {https://doi.org/10.1103/physrevlett.100.236801} {\bibfield  {journal}
  {\bibinfo  {journal} {Phys. Rev. Lett.}\ }\textbf {\bibinfo {volume} {100}},\
  \bibinfo {pages} {236801} (\bibinfo {year} {2008})}\BibitemShut {NoStop}%
\bibitem [{\citenamefont {Betancur-Ocampo}(2018)}]{Betancur-Ocampo2018}%
  \BibitemOpen
  \bibfield  {author} {\bibinfo {author} {\bibfnamefont {Y.}~\bibnamefont
  {Betancur-Ocampo}},\ }\bibfield  {title} {\bibinfo {title} {Partial positive
  refraction in asymmetric veselago lenses of uniaxially strained graphene},\
  }\href {https://doi.org/10.1103/physrevb.98.205421} {\bibfield  {journal}
  {\bibinfo  {journal} {Phys. Rev. B}\ }\textbf {\bibinfo {volume} {98}},\
  \bibinfo {pages} {205421} (\bibinfo {year} {2018})}\BibitemShut {NoStop}%
\bibitem [{\citenamefont {Betancur-Ocampo}\ \emph {et~al.}(2019)\citenamefont
  {Betancur-Ocampo}, \citenamefont {Leyvraz},\ and\ \citenamefont
  {Stegmann}}]{Betancur-Ocampo2019}%
  \BibitemOpen
  \bibfield  {author} {\bibinfo {author} {\bibfnamefont {Y.}~\bibnamefont
  {Betancur-Ocampo}}, \bibinfo {author} {\bibfnamefont {F.}~\bibnamefont
  {Leyvraz}},\ and\ \bibinfo {author} {\bibfnamefont {T.}~\bibnamefont
  {Stegmann}},\ }\bibfield  {title} {\bibinfo {title} {Electron optics in
  phosphorene pn junctions: Negative reflection and anti-super-klein
  tunneling},\ }\href {https://doi.org/10.1021/acs.nanolett.9b02720} {\bibfield
   {journal} {\bibinfo  {journal} {Nano Lett.}\ }\textbf {\bibinfo {volume}
  {19}},\ \bibinfo {pages} {7760} (\bibinfo {year} {2019})}\BibitemShut
  {NoStop}%
\end{thebibliography}%

\end{document}